\documentclass[aps,prd,twocolumn,superscriptaddress,nofootinbib,floatfix]{revtex4-2}

\usepackage[T1]{fontenc}
\usepackage[utf8]{inputenc}
\usepackage{lmodern}
\usepackage[reqno]{amsmath}
\usepackage{bm,amssymb,mathtools}
\usepackage{graphicx}
\usepackage{booktabs}
\usepackage{xcolor}
\usepackage{tikz}
\usetikzlibrary{calc}
\usepackage{pgfplots}
\pgfplotsset{compat=1.18}
\usepackage[unicode=true,hypertexnames=false]{hyperref}
\hypersetup{colorlinks=true,linkcolor=blue,citecolor=blue,urlcolor=blue}

\newcommand{\e}{\epsilon}

\usepackage{morefloats}

\begin{document}

\title{Quark Mixing from a Lattice Flavon Model: A Four-Magnitude Parameterization}

\author{Vernon Barger}
\affiliation{Department of Physics, University of Wisconsin--Madison, Madison, Wisconsin 53706, USA}

\date{July 26, 2026}

\begin{abstract}
We show that quark weak mixing follows from the same one-flavon, three-messenger Froggatt--Nielsen construction, with a single hierarchy parameter $B$ (with $\epsilon\equiv 1/B$) and a rational-exponent ``$B$-lattice,'' that organizes the fermion mass spectrum in companion work. The lattice exponent structure, together with a four-phase multi-messenger coefficient model, accounts for the four CKM magnitudes and, through the Fritzsch--Xing reconstruction, the full $3\times 3$ CKM matrix and the CP-violating Jarlskog invariant, providing a consistency test of the single-$B$ hypothesis; a benchmark Yukawa pair on the lattice reproduces, under direct diagonalization, the light-quark masses to better than $5\%$, all nine CKM magnitudes to better than $1.2\%$, and $J$ within $0.3\sigma$ of the PDG global fit, with order-unity coefficients, and a $10^7$-configuration phase scan shows the CKM hierarchy to be generic once the quark masses are fit, while percent-level agreement across all four magnitudes is not. The near-maximal CP phase admits a geometric interpretation: equal-weight messenger interference fixes the phase at exactly $\pi/2$ for any individual phase values, and the fitted $\phi_{\rm FX}\simeq93^\circ$ measures the departure from this limit. The same lattice parameter that organizes quark mixing also fixes the charged-lepton mass ratio $m_\mu/m_\tau=c_\mu\,\e^{5/3}$, whose exponent $5/3=Q(L_2)+Q(e^c_2)$ is the second-generation lepton charge sum; this yields the integer-exponent relation $|V_{ub}|\simeq(m_\mu/m_\tau)^{2}$, which connects the smallest CKM magnitude to the charged-lepton spectrum with no additional free parameters and agrees with the measured value at about the $6\%$ level. An equivalent cross-sector form $|V_{ub}|\simeq\sin^3\theta_{13}$ ties the smallest CKM magnitude directly to the smallest PMNS magnitude; the two routes agree at the few-percent level, providing complementary determinations of the same ninths-lattice quantity from the quark and lepton sectors.
\end{abstract}

\maketitle

\section{Introduction and Scope}

We present the third paper in a flavon/Froggatt--Nielsen~\cite{FN1979} trilogy in which a one-flavon plus three-messenger-chain model generates a single hierarchy
parameter $B$ (with $\epsilon \equiv 1/B$) and a rational-exponent ``$B$-lattice'' that organize
fermion Yukawa textures.
Building on the mass fits established in the first two papers~\cite{Barger2026bfn,Barger2026bfnb}, we translate the lattice
structure into predictions for weak mixing in the quark sector.

A minimal three-angle plus one-phase Cabibbo--Kobayashi--Maskawa (CKM)~\cite{Cabibbo1963,KM1973} parametrization makes
the Cabibbo angle an interference between the up- and down-sector $(1,2)$ rotations, while
CP violation is governed by the same hierarchy through a compact Jarlskog invariant~\cite{Jarlskog}.
At a benchmark point motivated by the lattice, the resulting CKM magnitudes and Wolfenstein
parameters~\cite{Wolfenstein} agree with the Particle Data Group global fit~\cite{PDGCKM2024} to
better than $0.3\sigma$ and $0.5\sigma$, respectively.

We further exhibit ratio relations in which the leading $B$-power
dependence of CKM elements is isolated, providing multiple
consistency checks of the hierarchy parameter from weak-mixing data.
These ratios are insensitive to the order-one coefficients at the level
of the $B$-power counting but are not strictly coefficient-independent;
the role of the multi-messenger coefficient structure is
quantified in Sec.~\ref{sec:coefficient-sensitivity}.
Throughout, we keep three levels of claim distinct: the lattice fixes
the rational exponents; the benchmark of
Sec.~\ref{sec:cp-phase-lattice} demonstrates that an order-unity
coefficient realization exists; and no claim of uniqueness of that
realization is made.
We restrict attention to the quark sector and CKM mixing,
deferring lepton mixing~\cite{Pontecorvo,MNS} to a separate analysis.

\section{Single-\texorpdfstring{$B$}{B} Lattice Framework}

The starting point is a Froggatt--Nielsen (FN) construction~\cite{FN1979,Leurer1992,Leurer1993} in which all fermion mass hierarchies
arise from powers of a single small parameter $\epsilon \equiv 1/B$.
The effective Yukawa matrices may be written schematically as
\begin{equation}
Y_f \;=\; C_f \circ \epsilon^{\,p_f},
\end{equation}
where $p_f$ is a matrix of rational exponents fixed by the $B$-lattice and $C_f$ is a matrix
of complex order-one coefficients.
In the standard horizontal-symmetry construction, small integer charge differences under a single $U(1)$ horizontal symmetry
generate the observed quark and lepton mass and mixing hierarchies via
$y_{ij}\sim \epsilon^{|q_i+q_j|}$.
The $B$-lattice extends this program by promoting the horizontal symmetry to a discrete
$\mathbb{Z}_N$ gauge symmetry (Sec.~\ref{sec:UV-rational}), so that the charge
differences themselves are quantized in units of $1/N$, generating the rational-exponent
texture used throughout this paper.
The symbol ``$\circ$'' denotes element-wise multiplication. The frequent restriction to integer FN powers in the literature reflects implicit assumptions of a single abelian symmetry with minimal messenger structure; once these assumptions are relaxed to allow discrete gauge symmetries and lattice charge assignments, rational exponents arise naturally. As shown by Krauss–Wilczek~\cite{Krauss:1988zc}, Banks–Dine~\cite{BanksDine:1991xc}, and Ibáñez–Ross~\cite{IbanezRoss1991}, this integer prejudice is not justified. Discrete gauge symmetries can support fractional effective charges, and anomaly cancellation constrains combinations, not individual integers.

The rational-exponent structure encodes the family hierarchy and determines the dominant
scaling of masses and mixing angles, while the coefficients $C_f$ control subleading numerical
factors and phases.
Once the exponent matrices are fixed, the framework is predictive: entire classes
of observables depend only on powers of $B$, with limited sensitivity to the detailed coefficient
choices.

Diagonalization of the Yukawa matrices proceeds via bi-unitary transformations,
\begin{equation}
Y_f^{\mathrm{diag}} \;=\; U_{f_L}^\dagger\, Y_f\, U_{f_R},
\label{eq:bi-unitary}
\end{equation}
and the CKM matrix is given by
\begin{equation}
V_{\mathrm{CKM}} \;=\; U_{u_L}^\dagger U_{d_L}.
\end{equation}
In this framework, quark mixing arises from the controlled misalignment between the up-
and down-sector rotations dictated by the same underlying lattice.

\subsection{Hierarchy Parameter and Power Counting}

It is convenient to adopt a minimal three-angle plus one-phase parametrization of the CKM
matrix in which the Cabibbo angle appears as an interference between the $(1,2)$ rotations of
the up and down sectors.

The Fritzsch–Xing (FX) parameterization~\cite{Fritzsch} of the CKM matrix is
\begin{equation}
V_{\mathrm{CKM}} \;=\;
R_{12}(\theta_u)\,
R_{23}(\theta)\,
\mathrm{diag}(e^{-i\phi_{\rm FX}},\,1,\,1)\,
R_{12}^\dagger(\theta_d),
\label{eq:FXparam}
\end{equation}
where $\theta_u$ and $\theta_d$ arise from the up- and down-sector $(1,2)$ rotations, $\theta$
is the dominant $(2,3)$ mixing angle, and $\phi_{\rm FX}$ is the CP-violating phase.
This parameterization is exact; the compact relations used here arise from controlled small-angle expansions that are numerically accurate at current experimental precision.

Within the $B$-lattice framework, each angle is associated with a definite power of
$\epsilon$.
This implies a hierarchy
\begin{equation}
\theta_d\approx 2.3\,\theta_u \approx 4.8\,\theta
\end{equation}
and ensures that CP violation is governed by the same underlying parameter that controls the
fermion mass spectrum.
The resulting Jarlskog invariant takes a compact form proportional to a simple product of
$B$-powers, providing a transparent link between flavor hierarchies and CP violation.

In terms of $s_u\equiv\sin\theta_u$, $s_d\equiv\sin\theta_d$, and $s\equiv\sin\theta$ one obtains
\begin{align}
V_{us}&= s_u\,c_d\,c-c_u\,s_d\,e^{-i\phi_{\rm FX}},
&
V_{cb}&= s\,c_u,
&
V_{ub}&=s\,s_u,
\label{eq:ckm-elements-1}\\
V_{td}&=-s\,s_d,
&
V_{ts}&=-s\,c_d,
&
V_{tb}&=c,
\label{eq:ckm-elements-2}
\end{align}
with $c\equiv\cos\theta$ and similarly $c_{u,d}=\cos\theta_{u,d}$.

The Jarlskog invariant in this parameterization takes the compact form
(see Ref.~\cite{Fritzsch})
\begin{equation}
\label{eq:J-FX}
J = s_u c_u\, s_d c_d\, s^2 c\,\sin\phi_{\rm FX},
\end{equation}
making manifest that CP violation is unsuppressed in phase but is suppressed by the hierarchy of the mixing angles.

\subsection{Input Masses and Scale Choice}

Throughout this work we quote running quark masses evaluated at the scale $M_Z$.
The overall normalization of the Yukawa matrices is fixed by matching the third-family masses,
$m_t(M_Z)$ and $m_b(M_Z)$, while the lighter-family masses emerge as predictions controlled by
powers of the hierarchy parameter $\e = 1/B$ and order-one coefficients.
Unless stated otherwise, we use
\begin{equation}
m_t(M_Z)\simeq 169~\mathrm{GeV}, \qquad
m_b(M_Z)\simeq 2.79~\mathrm{GeV},
\end{equation}
consistent with standard renormalization-group evolution.

With this choice, the up- and down-type mass spectra take the schematic form
\begin{align}
m_u : m_c : m_t &\;\sim\; \e^{\,p_u} : \e^{\,p_c} : 1, \\
m_d : m_s : m_b &\;\sim\; \e^{\,p_d} : \e^{\,p_s} : 1 ,
\end{align}
where the rational exponents $p_f$ are fixed by the $B$-lattice.
The numerical agreement of the resulting masses with experiment provides an internal consistency
check on the lattice assignment used in the mixing analysis below.

\section{Four-Magnitude CKM Parameterization}

Rather than fitting individual CKM matrix elements independently, we take as input four
experimentally well-determined magnitudes,
\begin{equation}
|V_{us}|,\qquad |V_{cb}|,\qquad |V_{ub}|,\qquad |V_{td}|,
\end{equation}
and use them to fix the hierarchy parameter $B$ and the relative orientation of the up-
and down-sector rotations.
In the $B$-lattice framework, each of these magnitudes is associated with a definite power of
$\e$, up to order-one coefficients.

A key feature of this approach is that ratios of CKM elements often eliminate the unknown coefficients,
yielding relations that depend only on powers of $B$.
This allows multiple, independent determinations of $B$ from weak-mixing data alone,
providing a consistency check of the single-parameter hypothesis.

For convenience, the direct mapping from four CKM magnitudes to the FX parameters may be written as
\begin{equation}
\boxed{
\begin{aligned}
s_u &\simeq \left|\frac{V_{ub}}{V_{cb}}\right|, \qquad
s_d \simeq \left|\frac{V_{td}}{V_{cb}}\right|,\\[4pt]
s &\simeq \frac{|V_{cb}|}{\cos\theta_u}\;\simeq\;|V_{cb}|,\\[4pt]
\cos\phi_{\rm FX} &=
\frac{s_d^{\,2}+s_u^{\,2}-|V_{us}|^{2}}
     {2\,s_u s_d}\, ,
\end{aligned}}
\label{eq:four-mag-summary-app}
\end{equation}

Corrections are of relative order ${\cal O}(s^2)$ and are numerically negligible at current precision.
The correspondence between CKM inputs and FX parameters is summarized in Table~\ref{tab:four-mag-map-app}; supporting analytic identities and empirical regularities used in the reconstruction are collected in Appendix~\ref{app:CKM_mainresults}.

\begin{table}[htbp]
\centering
\caption{Mapping between CKM magnitudes and Fritzsch--Xing parameters in the four-magnitude reconstruction.}
\label{tab:four-mag-map-app}
\renewcommand{\arraystretch}{1.2}
\begin{tabular}{cc}
\hline
CKM input & FX parameter \\ \hline
$|V_{ub}|/|V_{cb}|$ & $s_u\simeq\sin\theta_u$ \\
$|V_{td}|/|V_{cb}|$ & $s_d\simeq\sin\theta_d$ \\
$|V_{cb}|$ & $s\simeq\sin\theta$ \\
$|V_{us}|$ & $\phi_{\rm FX}$ (via Cabibbo interference) \\
\hline
\end{tabular}
\end{table}

\subsection{\texorpdfstring{$V_{us}$}{Vus} from Light-Quark Mass Ratios}

The Cabibbo angle arises from an interference between the $(1,2)$ rotations of the up and down sectors.
From the FX parameterization~\eqref{eq:FXparam}, the full interference formula is
\begin{equation}
\label{cabibbo-interference}
|V_{us}| \;\simeq\; \bigl|\,s_u - s_d\,e^{-i\phi_{\rm FX}}\,\bigr|,
\end{equation}
with the $(1,2)$ mixing angles governed by light-quark mass ratios,
\begin{equation}
\theta_u \;\sim\; \sqrt{\frac{m_u}{m_c}},
\qquad
\theta_d \;\sim\; \sqrt{\frac{m_d}{m_s}}.
\end{equation}
Within the $B$-lattice, both ratios are fixed by powers of $\e$, implying that the Cabibbo angle
itself is controlled by the same hierarchy parameter that governs the quark mass spectrum.

The phase plays an essential quantitative role.
In the real-texture limit ($\phi_{\rm FX}=0$),
Eq.~\eqref{cabibbo-interference} reduces to the classic
Gatto--Sartori--Tonin relation~\cite{GST1968,Oakes1969}
$|V_{us}|\simeq|\theta_d-\theta_u|\approx 0.114$, which undershoots the
measured value by roughly a factor of two.
At the empirical near-maximal value $\phi_{\rm FX}\approx\pi/2$ (Sec.~\ref{subsec:fx-phase-pdg}), the two terms add in
quadrature,
\begin{equation}
|V_{us}| \;\simeq\; \sqrt{s_d^2+s_u^2} \;\approx\; 0.220,
\end{equation}
reproducing the observed value~\cite{PDGVudVus}.
This provides a direct and transparent link between light-quark mass ratios, the CP-violating
phase, and the dominant CKM mixing angle.

\subsection{\texorpdfstring{$V_{cb}$}{Vcb} and the \texorpdfstring{$B$}{B}-Power Relation}

In the lattice framework the dominant $(2,3)$ mixing angle is controlled by a single power of
the hierarchy parameter $\e = 1/B$.
To leading order one finds
\begin{equation}
|V_{cb}| \;\simeq\; \theta \;\sim\; \e^{\,p_{cb}},
\end{equation}
where the exponent $p_{cb}$ is fixed by the rational $B$-lattice assignment.
This relation is largely insensitive to order-one coefficients and therefore provides a
particularly clean determination of $B$ from weak-mixing data.

Using the experimental value of $|V_{cb}|$, the inferred value of $B$ agrees with that obtained
from the quark-mass spectrum, providing a nontrivial consistency check of the single-parameter
hypothesis.

\subsection{\texorpdfstring{$V_{ub}$}{Vub} and \texorpdfstring{$V_{td}$}{Vtd}: Subleading Structure}

The remaining small CKM elements, $V_{ub}$ and $V_{td}$, arise from products of the $(1,2)$ and
$(2,3)$ rotations and therefore probe subleading powers of $\e$.
Parametrically, one finds
\begin{equation}
|V_{ub}| \;\sim\; \theta\,\theta_u,
\qquad
|V_{td}| \;\sim\; \theta\,\theta_d,
\end{equation}
up to order-one coefficients.

These relations imply the ratios
\begin{equation}
\frac{|V_{ub}|}{|V_{cb}|} \;\sim\; \theta_u,
\qquad
\frac{|V_{td}|}{|V_{cb}|} \;\sim\; \theta_d,
\end{equation}
which depend only on powers of $B$.
The observed hierarchy between $|V_{ub}|$ and $|V_{td}|$ therefore traces directly to the
hierarchy between light-quark mass ratios in the up and down sectors.

\subsection{Summary of CKM Magnitude Relations}

Collecting the leading results, the CKM magnitudes satisfy
\begin{align}
|V_{us}| &\sim |s_u - s_d\,e^{-i\phi_{\rm FX}}|, \\
|V_{cb}| &\sim \e^{\,p_{cb}}, \\
|V_{ub}| &\sim \e^{\,p_{cb}}\,\theta_u, \\
|V_{td}| &\sim \e^{\,p_{cb}}\,\theta_d ,
\end{align}
with all angles determined by rational powers of the single parameter $\e = 1/B$.
The success of these relations in reproducing the observed CKM hierarchy is consistent with
a single lattice parameter governing both quark masses and mixings.

\subsection{Charged-Lepton Mass Parameterization of the CKM}
\label{subsec:lepton-mass-CKM}

The same expansion parameter $\e = 1/B$ that controls quark
mixing also organizes the charged-lepton mass spectrum.
Consistent with the quark sector, we evaluate the charged-lepton
masses as running ($\overline{\rm MS}$) masses at the scale $M_Z$,
$m_e(M_Z)\simeq 0.487~\mathrm{MeV}$,
$m_\mu(M_Z)\simeq 102.7~\mathrm{MeV}$, and
$m_\tau(M_Z)\simeq 1746~\mathrm{MeV}$~\cite{Xing:2007fb}, so that all
masses entering the cross-sector relations are defined at a common scale.
The muon-to-tau mass ratio sits on the ninths lattice at
\begin{equation}
\frac{m_\mu}{m_\tau} \;=\; c_\mu\,\e^{5/3},
\qquad c_\mu\simeq 0.96,
\label{eq:mu-tau-Comp}
\end{equation}
where the exponent $5/3$ is the charged-lepton
mass-ratio exponent $p^\ell_{22}$, equal to the
second-generation lepton compositeness depth (the sum
of the lepton-doublet and charged-lepton-singlet charges
of the underlying construction~\cite{Barger2026bfnb,Subconstituents}),
\begin{equation}
\tfrac{5}{3} \;=\; p^\ell_{22} \;=\; Q(L_2)+Q(e^c_2)
\;=\; \tfrac{1}{2}+\tfrac{7}{6},
\label{eq:53-charges}
\end{equation}
with $Q(L_3)=Q(e^c_3)=0$ fixing the third generation as
the reference. The universal lepton-chain internal
factor $\Delta^\ell_{\rm int}=7/9$ cancels in the ratio,
so it does not enter Eq.~\eqref{eq:53-charges} (the full
diagonal Yukawa exponent is $p^\ell_{22}+\Delta^\ell_{\rm int}=22/9$).
Inverting
Eq.~\eqref{eq:mu-tau-Comp} yields the
\emph{master identity}
\begin{equation}
\e \;\simeq\; \left(\frac{m_\mu}{m_\tau}\right)^{\!3/5},
\label{eq:master-Comp}
\end{equation}
in which the exponent $3/5 = 1/p^\ell_{22} = 1/[Q(L_2)+Q(e^c_2)]$ is the
inverse of the same charge sum. This allows every CKM
magnitude $|V_{ij}|\sim\e^{p_{ij}}$
to be re-expressed as a power of the
muon-to-tau mass ratio with exponent $k_{ij}=3p_{ij}/5$:
\begin{align}
|V_{us}| &\;\sim\; \left(m_\mu/m_\tau\right)^{8/15}
\!\simeq 0.221, \\[2pt]
|V_{cb}| &\;\sim\; \left(m_\mu/m_\tau\right)^{17/15}
\!\simeq 0.040, \\[2pt]
|V_{td}| &\;\sim\; \left(m_\mu/m_\tau\right)^{5/3}
\!\simeq 0.0089, \\[2pt]
|V_{ub}| &\;\sim\; \left(m_\mu/m_\tau\right)^{2}
\!\simeq 0.0035.
\label{eq:CKM-lepton-mass-Comp}
\end{align}
The $|V_{td}|$ exponent $5/3$ is the algebraic
prediction from the Wolfenstein-product relation
$|V_{td}|\sim|V_{us}||V_{cb}|$, which sums the
$\e$-exponents $\tfrac{8}{9}+\tfrac{17}{9}=\tfrac{25}{9}$
and yields the integer-over-integer
$k=3\!\cdot\!\tfrac{25}{9}/5 = \tfrac{5}{3}$.
A best-fit variant $|V_{td}|\sim\e^{17/6}
=\e^{25.5/9}$, differing by half a ninth and absorbing
the residual unitarity-triangle phase factor,
gives instead $k=\tfrac{17}{10}$ and matches the data
to better than $1\%$ in the $\e$ form (see Sec.~VI).
Both forms agree on the central qualitative point:
$|V_{td}|$ lies in the integer-over-integer ninths
lattice between $|V_{cb}|$ and $|V_{ub}|$.

The exponent for $|V_{ub}|$ collapses to the
\emph{integer} $2$ because
$\tfrac{3}{5}\!\cdot\!\tfrac{10}{3} = 2$; equivalently,
the quark-mixing exponent $p(V_{ub})=\tfrac{10}{3}$ is
exactly twice the second-generation lepton depth,
$\tfrac{10}{3}=2\,p^\ell_{22}=2\,[Q(L_2)+Q(e^c_2)]$, so that
$k=p(V_{ub})/p^\ell_{22}=2$. This yields
the cross-sector identity
\begin{equation}
\boxed{\;|V_{ub}|\;\simeq\;
\left(\frac{m_\mu}{m_\tau}\right)^{\!2}\;}
\label{eq:Vub-lepton-mass-Comp}
\end{equation}
Numerically, $(5.88\!\times\!10^{-2})^2
= 3.46\!\times\!10^{-3}$, about $6\%$ below the
measured value
$|V_{ub}| = 3.69\!\times\!10^{-3}$~\cite{PDGCKM2024}
(and $\sim 9\%$ below the PDG inclusive--exclusive average $3.82\times10^{-3}$, whose larger uncertainty reflects the known inclusive--exclusive tension).

\paragraph*{Cross-sector quark--lepton mixing identity.}
The integer-exponent identity above admits a
companion form that bypasses the charged-lepton masses
entirely and expresses $|V_{ub}|$ in terms of a
\emph{PMNS} observable.
The leptonic reactor angle follows the
hop-charge prediction~\cite{Subconstituents,LeptonLattice}
\begin{equation}
\sin\theta_{13}\;\sim\;
\left(\frac{m_\mu}{m_\tau}\right)^{\!2/3},
\label{eq:sin13-Comp}
\end{equation}
which, raised to the third power, equals
$(m_\mu/m_\tau)^{2}\simeq |V_{ub}|$.
Therefore
\begin{equation}
\boxed{\;|V_{ub}|\;\simeq\;\sin^3\theta_{13}\;}
\label{eq:Vub-sin13-Comp}
\end{equation}
Numerically, with the measured reactor angle
$\sin\theta_{13}\simeq 0.149$,
$(\sin\theta_{13})^3\simeq 3.3\!\times\!10^{-3}$,
agreeing with the measured
$|V_{ub}| = 3.69\!\times\!10^{-3}$ to within $11\%$.
Note that the two boxed forms coincide exactly only when
$\sin\theta_{13}$ takes its predicted lattice value
$(m_\mu/m_\tau)^{2/3}\simeq 0.151$; the measured
$\sin\theta_{13}\simeq 0.149$ differs from this by
$\sim 1.5\%$, so the charged-lepton-mass and PMNS routes
to $|V_{ub}|$ agree with each other at the few-percent
level rather than identically.
This cross-sector relation ties the smallest
quark-mixing magnitude directly to the smallest
lepton-mixing magnitude through a single integer
power, with both quantities now independently measured
at the few-percent level.
Equivalently, the same lattice exponent
$\e^{10/3}$ that fixes $|V_{ub}|$ also fixes
$\sin^3\theta_{13}$, so the two observables furnish
\emph{complementary determinations of the same
ninths-lattice quantity} from the quark and lepton
sectors.

These four identities extend the Cabibbo--$B$
relation $|V_{us}|^{9/8} = \e = 1/B$ across the
entire CKM through the single mass ratio
$m_\mu/m_\tau$, providing an experimentally
testable bridge between the quark and lepton
sectors built on nothing more than ninths-lattice
arithmetic.

\section{CP Phase and Lattice Structure}
\label{sec:cp-phase-lattice}

\subsection{Phase Conventions}

We adopt a convention in which CP violation is dominated by the down-type Yukawa sector.
The up-sector phases $\phi_u,\psi_u$ are fixed by the requirement that the singular values of $Y_u$ reproduce the up-quark mass spectrum; the residual physical CP violation then resides in the down-sector structure and is governed by the Jarlskog invariant.
This simplifies the interpretation of the CKM phase, while the physical CP content is convention-independent and carried by $J$.

\subsection{Two Independent Down-Sector Phases}

The down-type Yukawa matrix contains two independent physical phases, denoted $\phi_d$ and
$\psi_d$.
These phases enter the coefficient matrix $C^d$ and control the interference pattern responsible
for CP violation in the CKM matrix.

The physically meaningful constraint is on the Jarlskog invariant $J$, not on the individual
phases themselves.
Different choices of $(\phi_d,\psi_d)$ that reproduce the same $J$ are physically equivalent.

In this convention, the effective phase entering the
Cabibbo interference~\eqref{eq:cabibbo-interf} is determined entirely by
the down-sector phases.  Diagonalization of the Fritzsch-texture
down-type Yukawa matrix produces a left-handed rotation $U_{d_L}$ whose
$(1,2)$ component acquires a phase from the $(1,2)$ Yukawa entry
(carrying $\phi_d$), while the $(2,3)$ component acquires a phase from
the $(2,3)$ entry (carrying $\psi_d$).  Because $R_{23}(\theta)$ acts primarily in the $(2,3)$ sector (with $\cos\theta\approx 1$ in the $(1,2)$ block), the Cabibbo element is governed at leading order by the product $R_{12}(\theta_u)\,\mathrm{diag}(e^{-i\phi_{\rm FX}},1,1)\,R_{12}^\dagger(\theta_d)$, so the up- and down-sector $(1,2)$ rotations interfere with a relative phase, producing the interference structure 
$e^{i\phi_d}-e^{i\psi_d}$; the geometry of this difference, developed in
Sec.~\ref{subsec:geometric-origin}, places the FX phase at $\pi/2$ in the
equal-weight limit
\begin{align}
\phi_{\rm FX}
\;&\equiv\;
\arg\!\left(e^{i\phi_d}-e^{i\psi_d}\right) \\
\;&\approx\;\frac{\pi}{2}
\label{eq:phiFX-def}
\end{align}

The phase $\phi_{\rm FX}$ arises from the interference of the two complex
rotations and depends only on the relative phase $\phi_d-\psi_d$; the mean
phase $(\phi_d+\psi_d)/2$ is removable by quark-field rephasing.

A clarification on convention dependence is in order. The statement
that $\phi_{\rm FX}$ is controlled by the down-sector phase difference
alone is specific to the convention adopted here, in which the
up-sector phases are absorbed into the field definitions. The physical
CP-violating content of the CKM matrix is not a property of any single
sector: it is the rephasing-invariant Jarlskog determinant $J$, which
in a basis with both sectors complex receives comparable contributions
from the up- and down-sector rotations. We have verified that, upon
forward diagonalization of the benchmark Yukawa matrices, $J$ is of the
same order whether the CP phases are carried by the up or the down
sector, as expected for a convention-independent quantity. The
down-sector $\phi_{\rm FX}$ should therefore be read as a convenient
parameterization of $J$ within the chosen convention, not as a claim
that CP violation originates physically in the down sector. The CKM
results of this paper (Tables~\ref{tab:CKM_Bpowers}
and~\ref{tab:CKM_full3x3}, and the value of $J$) are obtained from the
four-magnitude Fritzsch--Xing reconstruction, which takes the measured
magnitudes as input and is manifestly rephasing invariant, and are
therefore independent of this convention choice; the benchmark Yukawa
matrices reproduce the same magnitudes and $J$ under direct
diagonalization (Table~\ref{tab:benchmark-forward}), providing an
independent check.

The retuned factorized shifts and the four phases of the Yukawa benchmark are determined jointly by a fit to the
quark masses and CKM magnitudes, while maintaining the FX interference phase at a near-maximal value. The CKM
verification above is logically orthogonal: the four-magnitude $B$-power relations fix the CKM hierarchy and, together
with $\phi_{\rm FX}$, the CP-violating scale, while discrete retunes of the factorized shifts control the detailed
quark-mass spectrum without altering the leading CKM power counting.

A benchmark choice of phases (in radians) for which the singular values of $Y_u$ and $Y_d$ reproduce the quark masses and the diagonalization of the same matrices reproduces the CKM magnitudes and the Jarlskog invariant [Table~\ref{tab:benchmark-forward}] is
\begin{equation}
\begin{aligned}
  \phi_u&\simeq 3.4944, &\qquad \psi_u&\simeq 0.5941,\\
  \phi_d&\simeq 2.3174, &\qquad \psi_d&\simeq 3.6462.
\end{aligned}
\label{eq:app-phases}
\end{equation}

With the shift vectors of Eq.~\eqref{eq:shift-vectors} and the phases of
Eq.~\eqref{eq:app-phases}, the Yukawa matrices of Eq.~\eqref{eq:fnmessenger}
are diagonalized numerically; the resulting quark masses and CKM magnitudes
are collected in Table~\ref{tab:benchmark-forward}. All four light-quark
masses are reproduced to better than $5\%$, all nine CKM magnitudes to better
than $1.2\%$, and the Jarlskog invariant to $0.1\%$ of the four-magnitude
reconstruction value, with the effective
coefficients remaining of order unity
($0.42\le|C^u_{ij}|\le 1.60$, $0.30\le|C^d_{ij}|\le 0.91$).
We note that the schematic equal-weight combination of
Eq.~\eqref{eq:phiFX-def}, evaluated at the phases of
Eq.~\eqref{eq:app-phases}, gives
$\arg(e^{i\phi_d}-e^{i\psi_d})\simeq 81^\circ$, twelve degrees below the
fitted $\phi_{\rm FX}\simeq 93^\circ$ of
Sec.~\ref{subsec:fx-phase-pdg}; the spread quantifies how far the
benchmark coefficients sit from the equal-weight limit of
Sec.~\ref{subsec:geometric-origin}, in which the schematic combination
would be exact.
The reconstruction of Sec.~\ref{sec:CKM_Bscaling}, which takes the four
measured magnitudes as input, and this direct diagonalization of the
benchmark Yukawa matrices are therefore mutually consistent.

We emphasize what this benchmark does and does not establish. It is a
specific point in the space of shift vectors and phases, located
numerically; it demonstrates that the lattice exponent structure
\emph{admits} a simultaneous description of the quark masses, the CKM
magnitudes, and $J$ with order-unity coefficients, but it does not
constitute a prediction of the mixing pattern from the mass spectrum.
The two statements are compatible: the Monte Carlo of
Sec.~\ref{sec:coefficient-sensitivity} shows that mass-consistent
phase configurations reproduce the CKM hierarchy generically at the
factor-of-two level, while percent-level agreement across all four
magnitudes, together with the near-maximal interference phase, singles
out the benchmark point. The uniqueness of the benchmark has not been assessed here. The four
phases must be specified to about $10^{-2}$~rad: displacing the most
sensitive phase, $\phi_u$, by $0.01$~rad shifts individual masses and
CKM magnitudes by up to $3\%$.

\begin{table}[!tbp]
\caption{Forward diagonalization of the benchmark Yukawa matrices
[Eqs.~\eqref{eq:fnmessenger}, \eqref{eq:shift-vectors},
\eqref{eq:app-phases}], compared with data. Masses are
$\overline{\rm MS}$ at $M_Z$, normalized to $m_t=169$~GeV and
$m_b=2.794$~GeV; CKM magnitudes are compared with the PDG global
fit~\cite{PDGCKM2024}. The $J$ row is compared with the value implied by
the four input magnitudes and the measured phase,
$J=3.08\times10^{-5}$, itself consistent at $0.3\sigma$ with the PDG
global-fit $J=\left(3.12^{+0.13}_{-0.12}\right)\times10^{-5}$.
Row and column unitarity hold to $10^{-15}$.}
\label{tab:benchmark-forward}
\begin{tabular}{lccc}
\toprule
Observable & Benchmark & Data & Ratio \\
\midrule
$m_u$ [MeV] & 1.11 & 1.11 & 1.00 \\
$m_c$ [MeV] & 632 & 629 & 1.00 \\
$m_d$ [MeV] & 2.69 & 2.82 & 0.95 \\
$m_s$ [MeV] & 55.4 & 55.7 & 1.00 \\
\midrule
$|V_{ud}|$ & 0.97492 & 0.97435 & 1.001 \\
$|V_{us}|$ & 0.22250 & 0.22500 & 0.989 \\
$|V_{ub}|$ & 0.00371 & 0.00369 & 1.004 \\
$|V_{cd}|$ & 0.22237 & 0.22486 & 0.989 \\
$|V_{cs}|$ & 0.97406 & 0.97349 & 1.001 \\
$|V_{cb}|$ & 0.04202 & 0.04182 & 1.005 \\
$|V_{td}|$ & 0.00854 & 0.00857 & 0.997 \\
$|V_{ts}|$ & 0.04131 & 0.04110 & 1.005 \\
$|V_{tb}|$ & 0.99911 & 0.99912 & 1.000 \\
\midrule
$J$ & $3.08\times10^{-5}$ & $3.08\times10^{-5}$ & 1.001 \\
\bottomrule
\end{tabular}
\end{table}

\subsection{Rational Shift Vectors}

To implement the lattice structure at the coefficient level, we introduce rational shift vectors $\Delta,\Delta^\prime$ that modify the exponents of selected Yukawa entries. These reproduce the observed quark masses and CKM magnitudes while preserving the
underlying lattice relations.

The shift vectors define corresponding matrices $\Delta_d$ and $\Delta_d'$ that enter the
down-type Yukawa matrix as
\begin{equation}
Y_d \;=\; C^d \circ \e^{\,p_d + \Delta_d + \Delta_d'} .
\end{equation}
The explicit form of these matrices encodes the effect of the shifts while maintaining rational
exponents throughout; the benchmark exponent and shift matrices are given in Appendix~\ref{app:messenger-model-parameters}.

\subsection{FX CP Phase and Its PDG Counterpart}
\label{subsec:fx-phase-pdg}

The effective CKM phase $\phi_{\rm FX}$ is defined through the interference of the
down-sector phases and governs CP violation through the Jarlskog invariant rather than
through direct identification with a particular CKM angle.

Inverting the four measured CKM magnitudes gives
\begin{equation}
\phi_{\rm FX} \;\approx\; 93^\circ \;\approx\; \frac{\pi}{2},
\end{equation}
close to the maximal-CP value $\pi/2$ that maximizes $J$ for fixed CKM magnitudes.
The exact FX inversion (Appendix~\ref{app:algebraic-dictionary}) gives
$\phi_{\rm FX}=93.1^\circ$; the leading-order formula of
Eq.~\eqref{eq:four-mag-summary-app} gives $91^\circ$. The two-degree spread
between them reflects the sensitivity of the angle near
$\cos\phi_{\rm FX}\approx 0$, where small approximation errors in the
magnitudes translate into degree-level shifts in the phase; we quote the
exact value throughout, consistent with Eq.~\eqref{eq:UT-angles} and
Table~\ref{tab:wolfenstein-compare-app}.
We stress that this near-maximality is an empirical feature taken from the data, not
a prediction of the lattice. The down-sector phases are fixed by the quark masses only
up to the residual freedom that determines $\phi_{\rm FX}$, and mass-consistent phase
choices yield $\phi_{\rm FX}$ over a wide range; the lattice/charge structure does not
single out $\pi/2$. An equal-weight normalization of the two interfering messenger
amplitudes would fix $\phi_{\rm FX}=\pi/2$ exactly
(Sec.~\ref{subsec:geometric-origin}); the fitted three-degree offset from $\pi/2$
then measures the departure from that limit. What the lattice does predict is the \emph{magnitude} of the
Jarlskog invariant through the exponent sum
$J\sim\e^{55/9}$ (the sum $8/9+17/9+10/3$ of the three mixing exponents); the CP
\emph{phase}, and hence the near-maximality of the observed CP violation, enters as the
experimental value of $\sin\delta$ rather than being derived.

The standard PDG phase $\delta_{\rm CKM}$ differs numerically from $\phi_{\rm FX}$
because $\delta_{\rm CKM}$ is a convention-dependent parameter and $\phi_{\rm FX}$ is a
parameterization-dependent quantity; the physically meaningful equality is for the
Jarlskog invariant, not for the phases themselves.

\subsection{A Geometric Origin for the Near-Maximal Phase}
\label{subsec:geometric-origin}

The near-maximality of $\phi_{\rm FX}$ admits a geometric interpretation
that sharpens the separation between prediction and input. For any pair
of phases, the equal-weight combination in Eq.~\eqref{eq:phiFX-def}
obeys the exact identity
\begin{equation}
e^{i\phi_d}-e^{i\psi_d}
\;=\;
2i\,\sin\!\left(\frac{\phi_d-\psi_d}{2}\right)
e^{\,i(\phi_d+\psi_d)/2},
\label{eq:chord-identity}
\end{equation}
so the difference of two unit phasors is orthogonal to their mean
direction: a chord of the unit circle is perpendicular to the radius
that bisects it (Fig.~\ref{fig:phasor-geometry}). The modulus
$2\,|\sin[(\phi_d-\psi_d)/2]|$ sets the strength of the interference,
while the argument is the mean phase plus exactly $\pi/2$. Since the
mean phase $(\phi_d+\psi_d)/2$ is removable by quark-field rephasing,
the equal-weight combination gives $\phi_{\rm FX}=\pi/2$ identically,
for any values of the individual phases.

\begin{figure}[tbp]
\centering
\begin{tikzpicture}[scale=1.85,>=latex]
  \def\half{38.07} 
  \def\rot{30}    
  \pgfmathsetmacro{\phid}{\rot+\half}
  \pgfmathsetmacro{\psid}{\rot-\half}
  \draw[gray!55] (0,0) circle (1);
  \draw[gray!45,->] (-1.25,0)--(1.45,0) node[below right,gray!75!black]{\footnotesize\textbf{Re}};
  \draw[gray!45,->] (0,-1.25)--(0,1.40) node[left,gray!75!black]{\footnotesize\textbf{Im}};
  \coordinate (O) at (0,0);
  \coordinate (P) at (\phid:1);
  \coordinate (Q) at (\psid:1);
  \coordinate (M) at ($ (P)!0.5!(Q) $);
  \draw[blue!80!black,line width=1.2pt,->] (O)--(P) node[above]{\footnotesize $e^{i\phi_d}$};
  \draw[blue!80!black,line width=1.2pt,->] (O)--(Q) node[below right]{\footnotesize $e^{i\psi_d}$};
  \draw[teal!55!black,dashed,thick,->] (O)--(\rot:1.14)
       node[right,teal!55!black]{\footnotesize bisector};
  \draw[red!80!black,line width=1.2pt,->] (Q)--(P);
  \node[red!75!black,anchor=south east] at ($(P)+(\phid+90:0.16)$)
       {\footnotesize $e^{i\phi_d}-e^{i\psi_d}$};
  \draw[black] ($(M)+(\rot+180:0.095)$)
            -- ($(M)+(\rot+180:0.095)+(\rot+90:0.095)$)
            -- ($(M)+(\rot+90:0.095)$);
  \draw[gray!70] (\psid:0.28) arc (\psid:\phid:0.28);
  \node[gray!55!black,anchor=west,inner sep=1pt] at (0.30,0.06){\footnotesize $\tfrac{\phi_d\!-\!\psi_d}{2}$};
  \node[red!70!black,anchor=east] (chordlen) at (94:0.42)
       {\footnotesize $2\sin\!\tfrac{\phi_d-\psi_d}{2}$};
  \draw[red!60,dashed,thin,shorten <=2pt,shorten >=1.5pt]
       (chordlen.east) -- ($(M)!0.4!(P)$);
\end{tikzpicture}
\caption{Geometric origin of the near-maximal phase $\phi_{\rm FX}$. The two
down-sector messenger amplitudes are represented as equal-weight unit
phasors $e^{i\phi_d}$ and $e^{i\psi_d}$ (blue). Their difference (red),
which carries the effective phase $\phi_{\rm FX}$, is perpendicular to
their bisector (teal dashed) by the identity~\eqref{eq:chord-identity}:
a chord of the unit circle is orthogonal to the radius bisecting it.
The orthogonality, hence $\phi_{\rm FX}=\pi/2$ after removal of the
mean phase, follows from the equal weights alone, independent of the
individual phase values, while the chord length
$2\sin[(\phi_d-\psi_d)/2]$ sets the interference strength. The
phase separation is drawn to scale at the benchmark value of
Eq.~\eqref{eq:app-phases}, $|\phi_d-\psi_d|=76.1^\circ$ (half-angle
$38.1^\circ$), giving chord length $2\sin 38.1^\circ\simeq 1.23$; the
overall orientation, removable by quark-field rephasing, is rotated
for clarity. In the benchmark
coefficient model the two amplitudes carry unequal
$\epsilon$-suppressions, and the fitted $\phi_{\rm FX}\simeq93^\circ$
measures the departure from this equal-weight limit.}
\label{fig:phasor-geometry}
\end{figure}

The operative condition is therefore not a phase alignment but a
normalization: $\phi_{\rm FX}=\pi/2$ is exact when, and only when, the
two interfering messenger amplitudes enter with equal weight. In the
explicit coefficient model of Eq.~\eqref{eq:fnmessenger} this limit is
not literally realized: the two phase-carrying terms of the $(1,2)$
down-sector entry carry the suppressions
$\e^{\Delta^d_{12}}=\e^{13/9}$ and $\e^{\Delta'^d_{12}}=\e^{10/9}$,
whose ratio $\e^{1/3}\simeq0.57$ measures the departure from equal
weight, and the benchmark coefficients span
$0.30\le|C^d_{ij}|\le0.91$. The data quantify the same departure from
the other side: the exact inversion gives
$\phi_{\rm FX}\simeq93.1^\circ$, three degrees from the equal-weight
value, while the schematic combination evaluated at the benchmark
phases of Eq.~\eqref{eq:app-phases} gives
$\arg(e^{i\phi_d}-e^{i\psi_d})\simeq81^\circ$, twelve degrees below the
fitted value. Both offsets measure the unequal weights and subleading
rotations that the equal-weight schematic omits.

Two consequences follow. First, because $\sin 93.1^\circ=0.9986$, the
observed $J$ lies within $0.15\%$ of the ceiling
$J_{\max}= s_u c_u\, s_d c_d\, s^2 c \simeq 3.08\times10^{-5}$ fixed by
the magnitudes alone through Eq.~\eqref{eq:J-FX}; the physically sharp
restatement of ``$\alpha$ near $90^\circ$'' is ``CP violation
near-maximal.'' Second, the apex angle obeys
$\alpha\simeq\phi_{\rm FX}-\arctan(s_u s_d)\simeq\phi_{\rm FX}-1.0^\circ$,
so the equal-weight limit predicts the triangle
$(\alpha,\beta,\gamma)\simeq(89.0^\circ,\,23.0^\circ,\,68.1^\circ)$,
while the fitted point gives Eq.~\eqref{eq:UT-angles}. The two
triangles differ only through the three-degree offset in
$\phi_{\rm FX}$; both are consistent with the direct determinations
$\gamma=(65.7\pm3.0)^\circ$ and
$\alpha=\left(84.1^{+4.5}_{-3.8}\right)^\circ$~\cite{PDGCKM2024}: the
fitted and equal-weight $\gamma$ lie at $-0.1\sigma$ and $+0.8\sigma$,
while the corresponding $\alpha$ values lie $1.8\sigma$ and
$1.1\sigma$ above the direct determination, whose central value also
sits below the global-fit apex.

We emphasize what the geometry does and does not establish. It does not
derive the phase from the lattice charges: the equal-weight condition
is a structural hypothesis about the messenger normalization, not a
consequence of the $\mathbb{Z}_9$ charge assignments, consistent with
the Monte Carlo of Sec.~\ref{sec:coefficient-sensitivity} in which
generic mass-consistent phase configurations spread $\phi_{\rm FX}$
over a wide range. A direct attempt to fix the phase from the discrete
charges fails as well: restricting the down-sector rotation phases to
$\mathbb{Z}_9$ values $\{0,2\pi/9,\dots\}$, the assignment that best
reproduces $J$ yields $\alpha\simeq68^\circ$, excluded at more than
three standard deviations by the measured angle, and no assignment
reproduces both observables. What the geometry does establish is a
sharp reformulation of the empirical input: the measured CKM matrix
sits within three degrees of the equal-weight limit of the messenger
construction, and the offset from $\pi/2$ is a measurement of the
weight asymmetry rather than an independent phase parameter.

\section{Benchmark Fit and Quantitative Results}

Using the four CKM magnitudes as input, we determine the effective phase $\phi_{\rm FX}$ that
reproduces the observed CP violation.
The resulting CKM matrix agrees with global-fit values within quoted uncertainties.

\subsection{Sharp Ratio Relations}
Even before specifying order-one coefficients, Eqs.~\eqref{eq:ckm-elements-1} and \eqref{eq:ckm-elements-2} imply
several clean relations, including
\begin{equation}
\setlength{\arraycolsep}{2pt}
\begin{array}{@{}l@{\hspace{0.8cm}}l@{}}
\displaystyle
\begin{aligned}
\frac{|V_{ub}|}{|V_{cb}|} &\simeq \tan\theta_u \\
&\approx 0.0886,
\end{aligned}
&
\displaystyle
\begin{aligned}
\frac{|V_{td}|}{|V_{ts}|} &\simeq \tan\theta_d \\
&\approx 0.2083,
\end{aligned}
\\[4pt]
\multicolumn{2}{@{}l@{}}{\displaystyle
\begin{aligned}
\frac{|V_{td}|}{|V_{ub}|} &\simeq \frac{s_d}{s_u} \\
&\approx 2.312,
\end{aligned}}
\end{array}
\label{eq:ratio-preds}
\end{equation}
evaluated at the $B$-scaling magnitudes of
Table~\ref{tab:CKM_Bpowers}.

\begin{figure*}[tbp]
\centering
\begin{tikzpicture}
\begin{axis}[
  width=0.7\textwidth,
  height=0.3\textwidth,
  ymin=0.90, ymax=1.10,
  ymajorgrids,
  symbolic x coords={r1,r2,r3},
  xtick=data,
  xticklabels={$|V_{ub}|/|V_{cb}|$,$|V_{td}|/|V_{ts}|$,$|V_{td}|/|V_{ub}|$},
  xticklabel style={font=\scriptsize, align=center},
  yticklabel style={font=\scriptsize},
  ylabel={PDG / Eq.~\eqref{eq:ratio-preds}},
  legend style={font=\scriptsize, at={(0.02,0.98)}, anchor=north west, draw=none, fill=none},
]
\addplot+[
  only marks,
  mark=*,
  error bars/.cd,
    y dir=both,
    y explicit,
] coordinates {
  (r1, 1.0073) += (0, 0.0416) -= (0, 0.0412)
  (r2, 1.0018) += (0, 0.0394) -= (0, 0.0379)
  (r3, 0.9945) += (0, 0.0457) -= (0, 0.0427)
};
\addlegendentry{PDG global fit}

\addplot+[mark=none, red, thick, line cap=round] coordinates {(r1,1) (r2,1) (r3,1)};
\addlegendentry{Eq.~\eqref{eq:ratio-preds}}

\end{axis}
\end{tikzpicture}
\caption{Comparison of the sharp FX ratio predictions in Eq.~\eqref{eq:ratio-preds} with the PDG
global-fit magnitudes of the CKM matrix. The plotted points show the PDG central values divided by
our Eq.~\eqref{eq:ratio-preds} predictions; error bars are obtained by propagating the PDG
(one-sigma) uncertainties in Eq.~(12.27) of the PDG CKM review \cite{PDGCKM2024}.}
\label{fig:ratio-pdg}
\end{figure*}
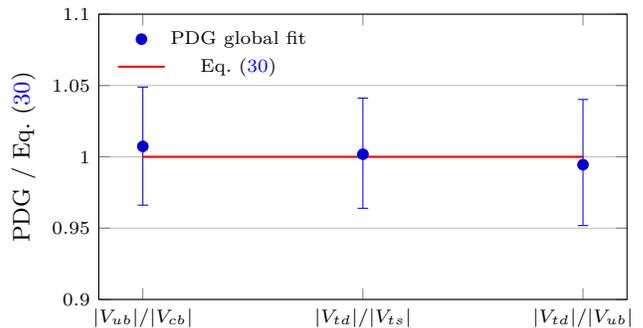

\noindent Figure~\ref{fig:ratio-pdg} displays these three ratios as (PDG)/(Eq.~\eqref{eq:ratio-preds}) with 1$\sigma$ uncertainties.

\begin{table}[htbp]
\centering
\caption{$B$ inferred from CKM magnitudes using the $B$-scaling relations and the experimental CKM determinations~\cite{PDGCKM2024}.}
\label{tab:B-from-CKM-app}
\renewcommand{\arraystretch}{1.12}
\begin{tabular}{lc}
\hline
Estimator & Value \\
\hline
$B_{us/cb}\equiv |V_{us}|/|V_{cb}|$ & 5.380201 \\
$B_{cb^3/ub^2}\equiv |V_{cb}|^3/|V_{ub}|^2$ & 5.371547 \\
$B_{ub^{1/2}/us^3}\equiv |V_{ub}|^{1/2}/|V_{us}|^3$ & 5.332927 \\
$B_{(td/ub)^2}\equiv (|V_{td}|/|V_{ub}|)^2$ & 5.393975 \\
\hline
$B_{V_{us}}\equiv |V_{us}|^{-9/8}$ & 5.355434 \\
$B_{V_{cb}}\equiv |V_{cb}|^{-9/17}$ & 5.368532 \\
$B_{V_{ub}}\equiv |V_{ub}|^{-3/10}$ & 5.368984 \\
$B_{V_{td}}\equiv |V_{td}|^{-6/17}$ & 5.364586 \\
\hline
\end{tabular}
\end{table}

\subsection{Direct Inversion of FX Angles}
\label{subsec:inversion}

A useful feature of the \mbox{Fritzsch--Xing} (FX) factorization is that the
\emph{magnitudes} of the measured CKM elements can be inverted directly
to obtain the underlying effective angles $(\theta_u,\theta_d,\theta)$
up to small ${\cal O}(s^2)$ corrections:
\begin{align}
\tan\theta_u &\simeq \left|\frac{V_{ub}}{V_{cb}}\right|\simeq B^{-13/9}, \nonumber\\
\tan\theta_d &\simeq \left|\frac{V_{td}}{V_{ts}}\right|\simeq B^{-17/18}, \nonumber\\
\sin\theta &\simeq \frac{|V_{cb}|}{\cos\theta_u} \simeq |V_{cb}|\simeq B^{-17/9}.
\label{eq:invert-angles}
\end{align}
The $B$-scaling values at $B=75/14$ are
\begin{equation}
\theta_u\simeq 5.09^\circ,\qquad \theta_d\simeq 11.6^\circ,\qquad \theta\simeq 2.40^\circ.
\end{equation}

From Eq.~\eqref{eq:J-FX}, the numerical value of the Jarlskog parameter is
\begin{equation}
J \simeq 3.1 \times 10^{-5}\, \sin{\phi_{\rm FX}}
\;\simeq\; c_J\, B^{-37/6}\, \sin{\phi_{\rm FX}},
\end{equation}
where the leading $B$-power $B^{-37/6}\simeq 3.2\times 10^{-5}$ captures the
hierarchy scaling and $c_J\simeq 0.97$ is an order-one factor from the
$\cos\theta_{u,d}$ corrections.
At the fitted phase $\phi_{\rm FX}\simeq 93^\circ$ (for which $\sin\phi_{\rm FX}\simeq 1.00$), this yields
\begin{equation}
J \;\simeq\; 3.1\times 10^{-5},
\end{equation}
consistent at $0.3\sigma$ with the PDG global-fit value $J=\left(3.12^{+0.13}_{-0.12}\right)\times 10^{-5}$~\cite{PDGCKM2024}.

\label{app:B-figure}
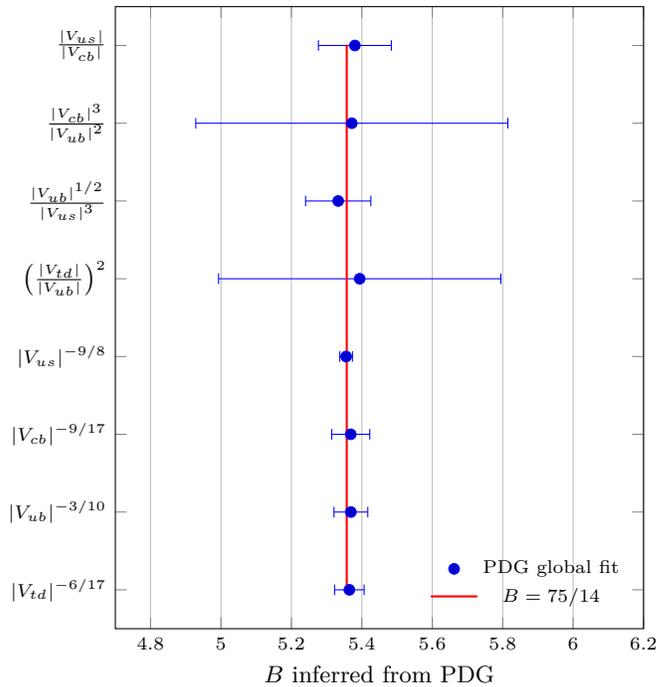
\begin{figure*}[tbp]
\centering
\begin{tikzpicture}
\begin{axis}[
  width=0.7\textwidth,
  height=0.42\textwidth,
  xmin=4.7, xmax=6.2,
  ymin=-0.5, ymax=7.5,
  xmajorgrids,
  ytick={0,1,2,3,4,5,6,7},
  yticklabels={$|V_{td}|^{-6/17}$,$|V_{ub}|^{-3/10}$,$|V_{cb}|^{-9/17}$,$|V_{us}|^{-9/8}$,$\left(\frac{|V_{td}|}{|V_{ub}|}\right)^{2}$,$\frac{|V_{ub}|^{1/2}}{|V_{us}|^{3}}$,$\frac{|V_{cb}|^{3}}{|V_{ub}|^{2}}$,$\frac{|V_{us}|}{|V_{cb}|}$},
  yticklabel style={font=\scriptsize},
  xticklabel style={font=\scriptsize},
  xlabel={$B$ inferred from PDG},
  legend style={font=\scriptsize, at={(0.98,0.02)}, anchor=south east, draw=none, fill=none},
]
\addplot+[
  only marks,
  mark=*,
  error bars/.cd,
    x dir=both,
    x explicit,
] coordinates {
  (5.380201, 7) += (0.103525, 0) -= (0.103525, 0)
  (5.371547, 6) += (0.443179, 0) -= (0.443179, 0)
  (5.332927, 5) += (0.092671, 0) -= (0.092671, 0)
  (5.393975, 4) += (0.400781, 0) -= (0.400781, 0)
  (5.355434, 3) += (0.017941, 0) -= (0.017941, 0)
  (5.368532, 2) += (0.054030, 0) -= (0.054030, 0)
  (5.368984, 1) += (0.048015, 0) -= (0.048015, 0)
  (5.364586, 0) += (0.041977, 0) -= (0.041977, 0)
};
\addlegendentry{PDG global fit}

\addplot+[mark=none, red, thick, line cap=round] coordinates {(5.3571,0) (5.3571,7)};
\addlegendentry{$B=75/14$}

\end{axis}
\end{tikzpicture}
\caption{$B$ inferred from CKM magnitudes using the PDG Standard-Model global-fit values (unitarity imposed), with $1\sigma$ error bars from standard propagation.
The vertical reference line is drawn at $B=75/14\simeq 5.357$.}
\label{fig:B-estimators}
\end{figure*}

\subsection{\texorpdfstring{$B$}{B}-identities from CKM ratios}
\label{app:B-identities}

Several CKM ratios provide direct determinations of $B$ with weak sensitivity to unknown
${\cal O}(1)$ coefficient magnitudes. A representative set is
\begin{equation}
\setlength{\arraycolsep}{3pt}
\begin{array}{@{}l@{\hspace{0.9cm}}l@{}}
\displaystyle \frac{|V_{us}|}{|V_{cb}|} = B, &
\displaystyle \frac{|V_{cb}|^{3}}{|V_{ub}|^{2}} = B, \\[8pt]
\displaystyle \frac{|V_{ub}|^{1/2}}{|V_{us}|^{3}} = B, &
\displaystyle \left(\frac{|V_{td}|}{|V_{ub}|}\right)^{2} = B.
\end{array}
\label{eq:B-identities-app}
\end{equation}

The agreement between quark masses, CKM magnitudes, and CP violation provides a nontrivial
consistency check of the single-$B$ lattice hypothesis;
the numerical values of $B$ inferred from each estimator are collected in Table~\ref{tab:B-from-CKM-app} and displayed graphically in Figure~\ref{fig:B-estimators}.
No additional small parameters are required.

It is important to clarify the role of the order-one coefficients
in these ratio tests.
At the level of $B$-power counting, the ratio relations in
Eq.~\eqref{eq:B-identities-app} and the FX angle ratios in
Eq.~\eqref{eq:ratio-preds} isolate the leading exponent
dependence on $B$, which is fixed by the rational lattice alone.
The physical CKM elements also depend on the
$\mathcal{O}(1)$ coefficient matrices $C_f$ through the
Yukawa diagonalization, so the $B$-estimators are not strictly
coefficient-independent.
However, as shown in Sec.~\ref{sec:coefficient-sensitivity},
the multi-messenger phases are not free parameters
specific to the CKM fit: they are constrained by the quark mass
spectrum, and a Monte Carlo scan shows that the exponent structure
fixes the CKM hierarchy robustly, with mass-consistent phase
configurations reproducing the observed pattern at the factor-of-two
level with high probability. The power-counting content of the
estimators is therefore robust against the coefficient freedom, and
the percent-level agreement of Table~\ref{tab:B-from-CKM-app}
constitutes the additional, non-trivial test beyond the mass fit.
At the benchmark point (with the multi-messenger phases fit
jointly to the quark masses and CKM magnitudes), all eight
estimators in Table~\ref{tab:B-from-CKM-app} agree to within
$\pm 1\%$, and all four CKM magnitudes are reproduced to
better than $0.3\sigma$.


\section{Numerical \texorpdfstring{$B$}{B}-Scaling of CKM Magnitudes}
\label{sec:CKM_Bscaling}

A central phenomenological consequence of the lattice--flavon framework
is that quark flavor hierarchies are organized by a single parameter
\begin{equation}
B \;\equiv\; \epsilon^{-1}, \qquad \epsilon\ll 1,
\end{equation}
with rational powers fixed by the lattice/messenger structure.
For CKM magnitudes we obtain simple $B$--power relations of the form
\begin{equation}
|V_{ij}| \;\sim\; B^{-p_{ij}}
\label{eq:Bpower_general}
\end{equation}
where the $p_{ij}$ values are given in Table~\ref{tab:CKM_Bpowers} for the benchmark $B$-value
\begin{equation}
B=\frac{75}{14}\simeq 5.357 .
\end{equation}

\subsection{Four Magnitudes as Rational B--Powers}

At leading order, the four CKM magnitudes are reproduced by the compact set of rational exponents in the second column of Table~\ref{tab:CKM_Bpowers}.

These relations capture the observed strong hierarchy
$|V_{us}|\gg |V_{cb}|\gg |V_{ub}|$ while tying it to a single organizing
parameter rather than independent small numbers.

\subsection{Numerical Illustration at \texorpdfstring{$B=75/14$}{B = 75/14}}

The resulting CKM magnitudes are given in the third column of Table~\ref{tab:CKM_Bpowers}. This indicates that a single $B$ with rational exponents can account for the four measured CKM magnitudes to good accuracy.

\begin{table}[htbp]
\centering
\caption{B--power values for CKM magnitudes at $B=75/14$.
Experimental values are from the PDG global fit (unitarity imposed)~\cite{PDGCKM2024}. The powers $p_{ij}$ in $|V_{ij}| = B^{-p_{ij}}$ are directly traceable to the $B^{-n}$ integer scaling of 2-over-2 masses~\cite{Barger2026bfnb}.}
\label{tab:CKM_Bpowers}
\begin{tabular}{lccc}
\toprule
Element & $B^{-p_{ij}}$ & $B$-scaling & Experiment \\
\midrule
$|V_{us}|$ & $B^{-8/9}$   & $0.2249$ & $0.22500\pm0.00067$ \\
$|V_{cb}|$ & $B^{-17/9}$  & $0.0420$ & $0.04182^{+0.00085}_{-0.00074}$ \\
$|V_{ub}|$ & $B^{-10/3}$  & $0.00372$ & $0.00369\pm0.00011$ \\
$|V_{td}|$ & $B^{-17/6}$  & $0.00860$ & $0.00857^{+0.00020}_{-0.00018}$ \\
\bottomrule
\end{tabular}
\end{table}

The quantitative agreement may be assessed via
\begin{equation}
\chi^2 \;=\; \sum_i \frac{\bigl(|V_{ij}|_{\rm pred} - |V_{ij}|_{\rm exp}\bigr)^2}{\sigma_i^2}
\;\simeq\; 0.17
\end{equation}
for the four magnitudes in Table~\ref{tab:CKM_Bpowers}, using the PDG global-fit
values~\cite{PDGCKM2024}, with all four elements agreeing to better than $0.3\sigma$.
We note that this small $\chi^2$ should be interpreted with
care.
Treating $B$ as the only free parameter
and the rational exponents as fixed a priori gives
$\chi^2/\mathrm{dof}\simeq 0.06$;
however, the multi-messenger phases
$(\phi_d,\psi_d)$ also contribute effective freedom
at the benchmark point (Sec.~\ref{sec:coefficient-sensitivity}).
A more robust characterization of the fit quality is the
\emph{consistency} of the multiple independent
$B$-determinations in Table~\ref{tab:B-from-CKM-app}:
the eight estimators span the range $5.33$--$5.39$
(a spread of $\pm 0.6\%$), consistent with
a single $B$ organizing the quark-mixing hierarchy.

\subsection{Comparison with the Wolfenstein Expansion}

\begin{table}[htbp]
\centering
\caption{Wolfenstein parameters (and the unitarity-triangle apex) compared with the PDG global fit and the PDG fit using only tree-level inputs.}
\label{tab:wolfenstein-compare-app}
\renewcommand{\arraystretch}{1.15}
\begin{tabular}{lcc}
\hline
 & This work & PDG global fit~\cite{PDGCKM2024} \\
\hline
$\lambda$ & 0.2249 & $0.22501\pm0.00068$ \\
$A$ & 0.8299 & $0.826^{+0.016}_{-0.015}$ \\
$\bar\rho$ & 0.1589 & $0.1591\pm0.0094$ \\
$\bar\eta$ & 0.3493 & $0.3523^{+0.0073}_{-0.0071}$ \\
\hline
\end{tabular}
\end{table}

It is instructive to compare this B--scaling organization with the
Wolfenstein parameterization (see Table~\ref{tab:wolfenstein-compare-app}), in which CKM hierarchies are expanded in
powers of the empirical parameter $\lambda\simeq |V_{us}|\approx 0.22$,
with additional coefficients $(A,\rho,\eta)$ encoding subleading structure
and CP violation. The associated unitarity-triangle angles in our flavon-messenger model are
\begin{equation}
(\alpha,\beta,\gamma)\approx (91.9^\circ,\ 22.6^\circ,\ 65.5^\circ).
\label{eq:UT-angles}
\end{equation}
In the present framework, the analogous ordering emerges from a
UV--motivated parameter $B=\epsilon^{-1}$ with rational exponents fixed by
lattice/messenger data: for example,
$|V_{us}|\sim B^{-8/9}$,
$|V_{cb}|\sim B^{-17/9}$,
and $|V_{ub}|\sim B^{-10/3}$ reproduce the familiar Wolfenstein pattern
while providing a dynamical origin for it.
Thus the Wolfenstein expansion may be viewed as an effective description
that can arise from an underlying lattice structure governed by a single
organizing parameter $B$; the explicit $B$-scalings of the Wolfenstein parameters are derived in Appendix~\ref{app:wolfenstein-scaling}.

\begin{table*}[tbp]
\centering
\footnotesize
\setlength{\tabcolsep}{2.5pt}
\caption{Key weak-mixing results and tests in the $B$-lattice framework.}
\label{tab:key-tests-app}
\renewcommand{\arraystretch}{1.12}
\begin{tabular}{p{0.32\linewidth}p{0.40\linewidth}p{0.20\linewidth}}
\hline
Observable & This work & Notes \\
\hline
Hierarchy pattern & $|V_{us}|\gg|V_{cb}|\gg|V_{ub}|,|V_{td}|$ & fixed by lattice \\
Improved powers & $|V_{us}|\sim\e^{8/9}$, $|V_{cb}|\sim\e^{17/9}$\newline $|V_{ub}|\sim\e^{10/3}$, $|V_{td}|\sim\e^{17/6}$ & see text \\
$B$ from ratios & $|V_{us}|/|V_{cb}|=B$,\newline $(|V_{td}|/|V_{ub}|)^2=B$, \dots & Eq.~\eqref{eq:B-identities-app} \\
Cabibbo interference & $|V_{us}|\simeq|s_u-s_d e^{-i\phi_{\rm FX}}|$ & Eq.~\eqref{eq:cabibbo-interf} \\
Phase relation & $\sin\phi_{\rm FX}\simeq \sin\delta_{\rm PDG}\,\frac{|V_{cb}||V_{us}|}{|V_{td}|}$ & Eq.~\eqref{eq:FX-PDG-relation-app} \\
Full CKM & all 9 $|V_{ij}|$ within $0.3\sigma$ & Table~\ref{tab:CKM_full3x3} \\
$|V_{ts}|$ consistency & $0.04128$ vs.\ PDG $0.04110$ & $+0.2\sigma$ \\
Jarlskog invariant & $J\simeq 3.1\times 10^{-5}$ & PDG: $\left(3.12^{+0.13}_{-0.12}\right)\!\times\!10^{-5}$ \\
Maximal-CP limit & $\phi_{\rm FX}=\pi/2$ exact for equal-weight messengers & Sec.~\ref{subsec:geometric-origin} \\
$B$ sensitivity & $B=5.345$--$5.373$ ($1\sigma$) & Fig.~\ref{fig:chi2-vs-B} \\
\hline
\end{tabular}
\end{table*}

\subsection{Full \texorpdfstring{$3\times 3$}{3x3} CKM Reconstruction}
\label{subsec:full-3x3}

Although only four magnitudes are used as input to fix the FX parameters, the
reconstruction determines the entire $3\times 3$ CKM matrix.
Table~\ref{tab:CKM_full3x3} compares all nine predicted magnitudes with the
PDG global-fit values~\cite{PDGCKM2024}.

\begin{table*}[tbp]
\centering
\caption{Full $3\times 3$ CKM magnitudes from the four-magnitude FX
reconstruction at $B=75/14$, compared with the PDG global fit (unitarity
imposed)~\cite{PDGCKM2024}. Elements marked with~($\star$) are the four
inputs; the remaining five are predictions.}
\label{tab:CKM_full3x3}
\renewcommand{\arraystretch}{1.15}
\begin{tabular}{lccc}
\toprule
Element & This work & PDG global fit & Pull \\
\midrule
$|V_{ud}|$        & $0.97438$ & $0.97435\pm0.00016$ & $+0.2\sigma$ \\
$|V_{us}|^\star$   & $0.22490$ & $0.22500\pm0.00067$ & $-0.1\sigma$ \\
$|V_{ub}|^\star$   & $0.00372$ & $0.00369\pm0.00011$ & $+0.3\sigma$ \\
\midrule
$|V_{cd}|$        & $0.22477$ & $0.22486\pm0.00067$ & $-0.1\sigma$ \\
$|V_{cs}|$        & $0.97351$ & $0.97349\pm0.00016$ & $+0.1\sigma$ \\
$|V_{cb}|^\star$   & $0.04200$ & $0.04182^{+0.00085}_{-0.00074}$ & $+0.2\sigma$ \\
\midrule
$|V_{td}|^\star$   & $0.00860$ & $0.00857^{+0.00020}_{-0.00018}$ & $+0.2\sigma$ \\
$|V_{ts}|$        & $0.04128$ & $0.04110^{+0.00083}_{-0.00072}$ & $+0.2\sigma$ \\
$|V_{tb}|$        & $0.99911$ & $0.99912\pm0.00025$ & $-0.0\sigma$ \\
\bottomrule
\end{tabular}
\end{table*}

Every element agrees with the PDG determination to better than $0.3\sigma$.
The total $\chi^2$ for all nine magnitudes is $\chi^2\simeq 0.3$; since five of
the nine follow from the four inputs by unitarity, this reflects the quality of
the four-magnitude reconstruction rather than nine independent tests.

Of the five predicted (non-input) elements, $|V_{ts}|$ provides a particularly
useful test.
In the FX parameterization $|V_{ts}|=s\,c_d$, which to leading $B$-power gives
$|V_{ts}|\sim B^{-17/9}$, the same scaling as $|V_{cb}|=s\,c_u$.
The two differ only through the ratio $c_d/c_u=\cos\theta_d/\cos\theta_u\simeq 0.983$,
yielding
\begin{equation}
|V_{ts}| \;\simeq\; 0.04128,
\end{equation}
compared with the PDG value $0.04110^{+0.00083}_{-0.00072}$ ($+0.2\sigma$).
Because $|V_{ts}|$ is not used as input, its agreement provides a useful
consistency check; within the FX factorization and three-generation
unitarity, however, $|V_{ts}|$ is fixed by the four inputs, so the test
probes the parameterization and the $B^{-17/9}$ scaling rather than an
independent lattice prediction.

\subsection{Unitarity Verification}
\label{subsec:unitarity}

Since the FX parameterization is an exact factorization of a unitary matrix,
the reconstructed CKM matrix satisfies row and column unitarity by construction.
We verify this explicitly at the benchmark point:
\begin{equation}
\sum_j |V_{ij}|^2 = 1, \qquad \sum_i |V_{ij}|^2 = 1,
\end{equation}
for all $i,j$, to numerical precision ($|1-\sum|<10^{-10}$).
This provides an internal consistency check on the four-magnitude reconstruction:
four inputs are sufficient to fix a unitary $3\times 3$ matrix (up to unphysical
phases), and the reconstruction is self-consistent.

\subsection{Sensitivity to \texorpdfstring{$B$}{B}}
\label{subsec:B-sensitivity}

\begin{figure*}[tbp]
\centering
\begin{tikzpicture}
\begin{axis}[
  width=0.7\textwidth,
  height=0.32\textwidth,
  xmin=5.28, xmax=5.45,
  ymin=0, ymax=30,
  xlabel={$B$},
  ylabel={$\chi^2$},
  xmajorgrids,
  ymajorgrids,
  xticklabel style={font=\scriptsize},
  yticklabel style={font=\scriptsize},
  legend style={font=\scriptsize, at={(0.80,0.96)}, anchor=north east, draw=none, fill=none},
]
\addplot[blue, thick, smooth] coordinates {
  (5.280, 28.95)
  (5.290, 22.01)
  (5.300, 16.05)
  (5.310, 11.06)
  (5.320, 7.02)
  (5.330, 3.92)
  (5.340, 1.75)
  (5.350, 0.49)
  (5.360, 0.14)
  (5.370, 0.67)
  (5.380, 2.09)
  (5.390, 4.37)
  (5.400, 7.51)
  (5.410, 11.48)
  (5.420, 16.29)
  (5.430, 21.92)
  (5.440, 28.36)
};
\addlegendentry{$\chi^2(B)$}

\addplot+[mark=none, red, thick, dashed, line cap=round]
  coordinates {(5.3571,0) (5.3571,30)};
\addlegendentry{$B=75/14$}

\addplot+[mark=none, black, dotted, line cap=round]
  coordinates {(5.28,1) (5.45,1)};
\addlegendentry{$\Delta\chi^2=1$}

\addplot+[mark=none, black!50, dotted, line cap=round]
  coordinates {(5.28,3.84) (5.45,3.84)};
\addlegendentry{$95\%$ CL}

\end{axis}
\end{tikzpicture}
\caption{$\chi^2$ for the four $B$-power CKM magnitude predictions in Table~\ref{tab:CKM_Bpowers} as a function
of $B$, using PDG global-fit values~\cite{PDGCKM2024}.
The minimum $\chi^2\simeq 0.13$ occurs at $B\simeq 5.359$, consistent with
$B=75/14\simeq 5.357$ (dashed red line).
The $\Delta\chi^2=1$ interval yields
$B=5.345$--$5.373$;
the $95\%$ CL interval ($\Delta\chi^2=3.84$) gives $B=5.330$--$5.388$.
The steep parabolic profile demonstrates that the four CKM magnitudes
tightly constrain $B$ to within $\pm 0.3\%$ at $1\sigma$.}
\label{fig:chi2-vs-B}
\end{figure*}
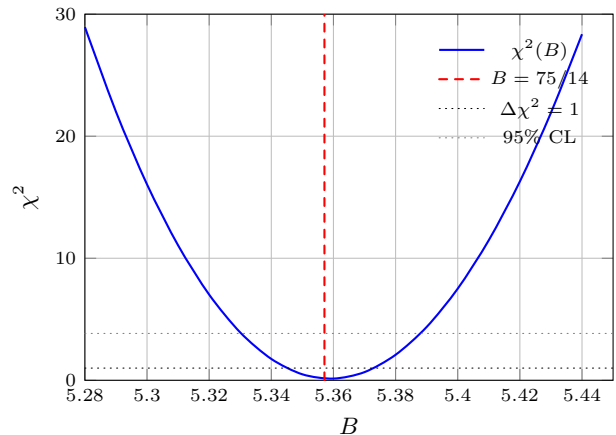

The sharpness of the $B$ determination can be assessed by scanning $\chi^2$ over
the hierarchy parameter while holding the rational exponents fixed.
Figure~\ref{fig:chi2-vs-B} shows the resulting $\chi^2$ profile for the four
CKM magnitudes in Table~\ref{tab:CKM_Bpowers}.
The minimum lies at $B\simeq 5.359$ with $\chi^2_{\min}\simeq 0.13$, and the
lattice value $B=75/14$ sits within $\Delta\chi^2<0.02$ of the minimum.
The $1\sigma$ range ($\Delta\chi^2=1$) is
\begin{equation}
B = 5.345\text{--}5.373 \qquad (1\sigma),
\end{equation}
corresponding to a fractional uncertainty of $\pm 0.3\%$.
The $\chi^2$ rises steeply outside this range: by $B=5.28$ or $5.44$ the fit is
excluded at the $5\sigma$ level.
This constraint from CKM data alone, consistent with the value extracted
independently from the quark mass spectrum~\cite{Barger2026bfn,Barger2026bfnb},
supports the single-$B$ lattice hypothesis.

\subsection{Coefficient Sensitivity of the $B$-Estimators}
\label{sec:coefficient-sensitivity}

A central question is the degree to which the
$B$-estimators of Eq.~\eqref{eq:B-identities-app}
depend on the $\mathcal{O}(1)$ coefficient matrices
$C_f$.
At the level of $B$-power counting, ratios such as
$|V_{us}|/|V_{cb}|\sim\epsilon^{-1}=B$ isolate the
leading exponent dependence.
However, the physical CKM elements also receive
$\mathcal{O}(1)$ prefactors from the Yukawa
diagonalization, so the estimators are not strictly
coefficient-independent.
We investigate this dependence with a Monte Carlo study.

\paragraph{Setup.}
We scan the four multi-messenger phases
$(\phi_u,\psi_u,\phi_d,\psi_d)$ uniformly over
$[0,2\pi)$, holding the lattice exponents and shift
matrices fixed.
For each realization, we construct the Yukawa matrices
via Eq.~\eqref{eq:fnmessenger}, diagonalize
numerically, and extract both the quark mass eigenvalues
and the CKM matrix.

\paragraph{Unconstrained scan.}
Even without any constraint, the exponent structure dominates the
mixing pattern: in an unconstrained sample of phase configurations,
roughly $30\%$ place $|V_{us}|$ in the observed range $[0.15,0.30]$,
while the individual $B$-estimators vary at the order-one level from
configuration to configuration. The estimators are therefore not
coefficient-free in the strict sense, but the underlying hierarchy is
set by the exponents rather than by the phases.

\paragraph{Mass-constrained scan.}
The multi-messenger phases are not free parameters of
the model: they are fixed by the requirement that
the Yukawa matrices reproduce the observed quark masses.
To test what the mass fit implies for mixing, we impose the constraint
that all four light-quark mass ratios
($m_u/m_t$, $m_c/m_t$, $m_d/m_b$, $m_s/m_b$)
lie within $\pm 30\%$ of their measured values.
Out of $10^7$ phase configurations, $11\,331$ ($0.11\%$) satisfy
this mass constraint.

Among these mass-consistent configurations, the CKM hierarchy is
reproduced generically. The central $68\%$ ranges are
$|V_{us}|=0.20$--$0.30$ (median $0.23$; observed $0.225$),
$|V_{cb}|=0.026$--$0.062$ (median $0.044$; observed $0.042$),
$|V_{ub}|=0.0037$--$0.0063$ (median $0.0049$; observed $0.0037$), and
$|V_{td}|=0.0087$--$0.014$ (median $0.011$; observed $0.0086$); the
Jarlskog invariant spans $|J|=(1.1$--$5.1)\times10^{-5}$ with median
$3.1\times10^{-5}$. In total, $84\%$ of mass-consistent configurations
place the Cabibbo angle in the observed range
$|V_{us}|\in[0.15,0.30]$, and about $80\%$ reproduce all four
magnitudes within a factor of two, with the observed values lying at
or near the medians of the scanned distributions. By contrast, only
$0.3\%$ reproduce all four magnitudes within $\pm10\%$, and the
scanned CP phase is unconstrained, with $|J|$ spanning a factor of
five. These fractions are stable under the gate definition: varying
the mass window between $\pm20\%$ and $\pm50\%$ moves the Cabibbo
fraction only between $79\%$ and $86\%$. An earlier version of this scan suffered from a
generation-ordering error in the CKM extraction, which inverted these
conclusions; the direct-diagonalization benchmark of
Table~\ref{tab:benchmark-forward} is unaffected
(Appendix~\ref{app:messenger-model-parameters}).

\paragraph{Interpretation.}
The correct characterization of the lattice predictions
is therefore as follows. The CKM \emph{hierarchy}, including the
order of magnitude of every mixing element and of $J$, is a robust
consequence of the rational exponent structure: it emerges at the
factor-of-two level for generic mass-consistent phases, and the
$B$-power content of the estimators in
Table~\ref{tab:B-from-CKM-app} is correspondingly insensitive to the
coefficient details. What the exponents alone do not fix are the
percent-level values of the magnitudes and the CP phase: only
$0.3\%$ of mass-consistent configurations match all four magnitudes
at the $\pm10\%$ level, and the phase spreads freely. The benchmark
point supplies exactly these two ingredients, percent-level
simultaneous agreement and a near-maximal interference phase, within
a system of eleven observables (four CKM magnitudes, six quark mass
ratios, and the CP phase) described by five continuous parameters (one hierarchy
parameter $B$ and four multi-messenger phases, with the discrete
shift matrices held fixed). The tight clustering
of the eight $B$-estimators in Table~\ref{tab:B-from-CKM-app}
(spread $<1\%$) reflects the same division: the clustering scale is
set by the exponents, the residual percent-level spread by the
coefficients.

\subsection{Renormalization Group Stability}
\label{sec:RGE-stability}

The $B$-lattice exponents of this paper, like the mass fits of the
companion work, are defined at $M_Z$, while an ultraviolet completion
lives at the flavor scale $\Lambda$. Between these scales the CKM
matrix evolves under the renormalization group (RG), so the matching
of the $M_Z$-scale lattice onto a UV construction must account for the
running. The one-loop structure is classic~\cite{Sasaki:1986jv,Babu:1987im}:
in the Standard Model the four elements connecting the third family,
$|V_{ub}|$, $|V_{cb}|$, $|V_{td}|$, and $|V_{ts}|$, evolve by a
\emph{common} multiplicative factor driven by $y_t^2$, increasing
toward the UV, while $|V_{us}|$ and the ratios
$|V_{ub}/V_{cb}|=\tan\theta_u$ and $|V_{td}/V_{ts}|=\tan\theta_d$
are RG-invariant to excellent accuracy, and $J$ scales as the square
of the common factor. Integrating the one-loop SM equations
numerically with the benchmark Yukawa matrices, the common factor is
$1.03$ at $\Lambda=10^{4}$~GeV, $1.06$ at $10^{6}$~GeV, $1.11$ at
$10^{12}$~GeV, and $1.13$ at $2\times10^{16}$~GeV, while $|V_{us}|$
and the two angle ratios shift by less than $0.02\%$ over the entire
range.

The consequences for the lattice are twofold. First, the exponent
\emph{differences} that fix the FX angles, and with them the entire
mixing hierarchy of Sec.~\ref{sec:coefficient-sensitivity}, are
scale-stable: $\tan\theta_u\simeq B^{-13/9}$ and
$\tan\theta_d\simeq B^{-17/18}$ hold at any scale at which they hold
at $M_Z$. Second, the overall $(2,3)$ power is not scale-stable: the
common RG factor is equivalent to shifting the $|V_{cb}|$ exponent
$17/9$ by $\ln\chi/\ln B\simeq 0.06$, about half a ninth, between
$M_Z$ and $10^{12}$~GeV. Estimators that mix $|V_{us}|$ with the
third-family elements, such as $|V_{us}|/|V_{cb}|$, correspondingly
drift by up to $\sim\!10\%$ over that range, so the tight
$\pm0.6\%$ clustering of Table~\ref{tab:B-from-CKM-app} is a
statement about the $M_Z$-scale CKM, where the lattice is defined. A
UV completion at scale $\Lambda$ must either match onto the
$M_Z$-scale exponents through the computable common factor above or
place the flavor scale low enough that the half-ninth shift is
negligible; we regard this matching as part of the UV construction
rather than of the present analysis.

\subsection{Rational Exponents and the UV Construction}
\label{sec:UV-rational}

Rational exponents $p_{ij}=n/9$ are not a physical
prediction unless the denominator and charge lattice
are fixed by an explicit ultraviolet construction.
Otherwise, one could always redefine
$\epsilon'\equiv\epsilon^{1/9}$ and express all
exponents as integers.
In the present framework, the ninths denominator is
fixed by the discrete $\mathbb{Z}_9$ gauge symmetry of the
companion chain construction of Ref.~\cite{Barger2026bfnb},
in which each effective Yukawa entry arises from unit-magnitude
messenger chains and the FN exponents are quantized in units of
$1/9$, each nearest-neighbor hop along the VLQ chain contributing
one charge quantum (see also~\cite{UFP}).
The rational exponent structure is therefore a
\emph{derived consequence} of the chain topology, not
an external ansatz.
The overdetermined fit (eleven observables, five parameters)
provides the empirical evidence that this specific
lattice is realized in nature, while the UV construction
provides the theoretical mechanism.

\section{Conclusion}

We have shown that quark masses, CKM magnitudes, and CP violation can be simultaneously
understood within a single-$B$ lattice framework in which rational exponents and controlled
phase structure replace arbitrary textures.
The $B$-scaling description traces both the magnitudes and CP-violating structure of the CKM
matrix to a common lattice origin, as summarized quantitatively in
Table~\ref{tab:key-tests-app}.
The four-magnitude parameterization provides a compact and practical tool for CKM analysis,
enabling direct extraction of the underlying mixing angles and phase from measured observables.

The full $3\times 3$ CKM reconstruction (Table~\ref{tab:CKM_full3x3}) yields all nine
magnitudes in agreement with the PDG global fit to better than $0.3\sigma$, with a
total $\chi^2\simeq 0.3$ for nine elements.
The Jarlskog invariant $J\simeq 3.1\times 10^{-5}$ agrees with the PDG global
fit at $0.3\sigma$,
and the $\chi^2$ profile in $B$ (Figure~\ref{fig:chi2-vs-B}) constrains the hierarchy
parameter to $\pm 0.3\%$ from CKM data alone, consistent with the value obtained
independently from quark masses.
Direct diagonalization of the benchmark Yukawa matrices reproduces the same
data (Table~\ref{tab:benchmark-forward}): the light-quark masses to better
than $5\%$, all nine CKM magnitudes to better than $1.2\%$, and $J$ to
$0.1\%$ of the reconstructed value, with every effective coefficient of order unity.

The CP phase acquires a geometric interpretation
(Sec.~\ref{subsec:geometric-origin}): the difference of two equal-weight
unit phasors is perpendicular to their bisector, so an equal-weight
normalization of the interfering messenger amplitudes fixes
$\phi_{\rm FX}=\pi/2$ exactly, for any values of the individual phases.
The lattice charges do not derive this normalization, and the fitted
$\phi_{\rm FX}\simeq 93^\circ$ shows the data sitting three degrees from
the equal-weight limit; the offset from $\pi/2$ is thereby recast as a
measurement of the messenger weight asymmetry rather than an
independent phase parameter, and the observed $J$ lies within $0.15\%$
of the ceiling fixed by the CKM magnitudes alone.

A central new result of this work is that the same lattice
parameter that governs quark mixing also organizes the
charged-lepton mass spectrum
(Sec.~\ref{subsec:lepton-mass-CKM}).
Inverting the muon--tau mass ratio
$m_\mu/m_\tau = c_\mu\,\e^{5/3}$, whose exponent
$5/3=p^\ell_{22}=Q(L_2)+Q(e^c_2)$ is the second-generation lepton
charge sum, to obtain the master
identity $\e\simeq(m_\mu/m_\tau)^{3/5}$ converts every
CKM magnitude into a power of the muon-to-tau mass
ratio: $|V_{us}|\sim(m_\mu/m_\tau)^{8/15}$,
$|V_{cb}|\sim(m_\mu/m_\tau)^{17/15}$,
$|V_{td}|\sim(m_\mu/m_\tau)^{5/3}$,
and the integer-exponent identity
\begin{equation}
|V_{ub}|\;\simeq\;(m_\mu/m_\tau)^{2}
\;=\;3.46\times 10^{-3},
\end{equation}
about $6\%$ below the measured value
$3.69\times 10^{-3}$. The integer exponent arises because
the quark-mixing exponent $p(V_{ub})=10/3$ is exactly
twice the second-generation lepton depth,
$10/3=2\,p^\ell_{22}=2[Q(L_2)+Q(e^c_2)]$.
The same lattice exponent $\e^{10/3}$ is also
predicted to control the PMNS reactor angle
through $\sin\theta_{13}\sim(m_\mu/m_\tau)^{2/3}$,
yielding the cross-sector
identity $|V_{ub}|\simeq\sin^3\theta_{13}\simeq
3.3\times 10^{-3}$ that ties the smallest CKM
and PMNS magnitudes to a common ninths-lattice
quantity~\cite{Subconstituents,LeptonLattice}.
These four relations bridge the quark and lepton
sectors through a single mass ratio with no fitted
parameter, extending the Cabibbo--$B$ identity
$|V_{us}|^{9/8}=1/B$ to the entire CKM and providing
a cross-sector consistency check of the single-$B$
hypothesis.

The success of these predictions, achieved with one hierarchy
parameter $B$ and four phases in the messenger coefficient
model fitted to an overdetermined system of eleven
observables, motivates further exploration of leptonic
mixing and possible ultraviolet completions, which we defer
to companion work~\cite{Barger2026bfnb,UFP,Subconstituents}.

\begin{acknowledgments}
VB gratefully acknowledges support from the
William F.\ Vilas Estate and from the
U.S.\ Department of Energy.
\end{acknowledgments}


\appendix

\section{Useful Analytical and Empirical Relations}
\label{app:CKM_mainresults}

This Appendix collects some useful analytic identities and empirical regularities in the
four-magnitude CKM reconstruction.

\subsection{Four-magnitude reconstruction in the Fritzsch--Xing form}
\label{app:four-mag-recon}

The CKM matrix in the minimal three-angle plus one-phase FX form may be parameterized
in terms of $s_u\equiv\sin\theta_u$, $s_d\equiv\sin\theta_d$, $s\equiv\sin\theta$ and
the single physical phase $\phi_{\rm FX}$.
To leading order in the small angles, the Cabibbo element is governed by interference,
\begin{equation}
|V_{us}|\simeq \left|\,s_u - s_d e^{-i\phi_{\rm FX}}\,\right|
\qquad(\text{to leading order in } s,s_{u,d}),
\label{eq:cabibbo-interf}
\end{equation}
while the remaining magnitudes reduce to the direct identifications
\begin{equation}
|V_{cb}|\simeq s,\qquad |V_{ub}|\simeq s\,s_u,\qquad |V_{td}|\simeq s\,s_d.
\label{eq:vcbubtd}
\end{equation}

\subsection{Approximate relation between \texorpdfstring{$\delta_{\rm PDG}$}{delta(PDG)} and \texorpdfstring{$\phi_{\rm FX}$}{phi(FX)}}
\label{app:phase-relation}

Equating the Jarlskog invariant $J$ across phase conventions yields a leading-order
relation connecting the PDG phase $\delta_{\rm PDG}$ to the FX phase $\phi_{\rm FX}$:
\begin{equation}
\sin\phi_{\rm FX}
\;\simeq\;
\sin\delta_{\rm PDG}\,
\frac{|V_{cb}|\,|V_{us}|}{|V_{td}|}\, ,
\label{eq:FX-PDG-relation-app}
\end{equation}
accurate to about $0.1\%$ in the sine (the cosine factors
$c_{12}c_{23}/(c_u c_d c)$ cancel to that level). Because both sines
are near unity, the inversion for the angles themselves is
ill-conditioned near maximal CP: a $0.1\%$ error in $\sin\phi_{\rm FX}$
corresponds to a degree-level shift in $\phi_{\rm FX}$, so the phase
should be extracted from the cosine form of
Eq.~\eqref{eq:cosphiFX-dict-app} or directly from $J$
[Eq.~\eqref{eq:sinphiFX-dict-app}] rather than from this relation.
This emphasizes that the physically meaningful equality is for $J$, not for the phases themselves.

\subsection{Empirical CKM magnitude relations and the unitarity question}
\label{app:empirical-ckm}

The observed quark mixing exhibits additional numerical regularities beyond the basic Wolfenstein hierarchy.
Grossman and Ruderman (GR) ~\cite{GrossmanRuderman} identified approximate relations among CKM magnitudes,
\begin{equation}
\setlength{\arraycolsep}{2pt}
\begin{array}{@{}l@{\hspace{1cm}}l@{}}
\displaystyle |V_{td}|^{2}\;\simeq\;|V_{cb}|^{3}, &
\displaystyle |V_{ub}|^{2}\,|V_{us}|\;\simeq\;|V_{cb}|^{4},
\end{array}
\label{eq:GR-relations-app}
\end{equation}
which hold at the few-percent level for PDG global-fit magnitudes.
We further note the additional empirical correlation
\begin{equation}
|V_{cb}|\,|V_{ub}|^{1/2}\;\simeq\;|V_{us}|^{4}.
\label{eq:new-empirical-app}
\end{equation}
The above empirical relations are useful for constructing multiple ``coefficient-clean'' estimators of the underlying hierarchy parameter $B$.

\paragraph{Relation to $B$-scaling.}
All three relations follow \emph{exactly} from the $B$-power assignments
in Table~\ref{tab:CKM_Bpowers}:
the first gives $B^{-17/3}=B^{-17/3}$, the second $B^{-68/9}=B^{-68/9}$,
and the third $B^{-32/9}=B^{-32/9}$.
The exponents match identically, not only to a few percent; the GR
relations are therefore built into the lattice structure rather than
being approximate numerical coincidences.

\paragraph{Are these relations consequences of unitarity?}
A $3\times 3$ unitary CKM matrix contains four physical parameters
(three angles and one phase).
Once four well-measured magnitudes are specified, unitarity fixes
the remaining elements, so one might ask whether the GR relations
are automatic.
They are not.
Expressed in Wolfenstein language, the first relation requires
$r_t^2/A=1$ and the second requires $r_u^2/(A^2\lambda)=1$, where
$r_{u,t}$ parametrize the unitarity-triangle sides.
These are nontrivial constraints on the Wolfenstein parameters
$(A,\bar\rho,\bar\eta)$ that are not imposed by unitarity alone;
a generic unitary CKM matrix with the measured $\lambda$ and $A$ need not
satisfy them.

\paragraph{The lattice as a dynamical origin.}
Grossman and Ruderman concluded that their relations may point to
``deeper structure'' but could not exclude an ${\cal O}(10\%)$ accident.
The $B$-lattice framework resolves this ambiguity: the rational exponents
$p_{ij}$ predict specific values of $A=\epsilon^{1/9}$,
$r_u=\epsilon^{5/9}$, and $r_t=\epsilon^{1/18}$
(Appendix~\ref{app:wolfenstein-scaling}),
which automatically enforce $r_t^2/A=1$ and $r_u^2/(A^2\lambda)=1$.
The GR relations therefore emerge as \emph{derived consequences} of the
lattice charge assignments rather than independent empirical
observations, and their percent-level accuracy in the data constitutes
additional evidence for the single-$B$ organizing principle.

\section{Flavor-Messenger Model Parameters}
\label{app:messenger-model-parameters}

\subsection{Effective quark Yukawa matrices}

We use the parameterization
\begin{equation}
  (Y_f)_{ij}=\epsilon^{\,p^{f}_{ij}}\Big[1+e^{i\phi_f}\epsilon^{\,\Delta^{f}_{ij}}+e^{i\psi_f}\epsilon^{\,\Delta^{\prime f}_{ij}}\Big]
  \label{eq:fnmessenger}
\end{equation}
where $f=u$ or $f=d$ and the exponent matrices are
\begin{equation}
\begin{aligned}
  p^{u}_{ij} &=
  {\setlength{\arraycolsep}{6pt}\renewcommand{\arraystretch}{1.15}
  \frac{1}{9}\begin{pmatrix}
    64 & 39 & 27\\
    55 & 30 & 18\\
    37 & 12  & 0
  \end{pmatrix}},\\[8pt]
  p^{d}_{ij} &=
  {\setlength{\arraycolsep}{6pt}\renewcommand{\arraystretch}{1.15}
  \frac{1}{9}\begin{pmatrix}
    37 & 30 & 27\\
    28 & 21 & 18\\
    10 & 3  & 0
  \end{pmatrix}}.
\end{aligned}
\label{eq:exponent-matrices}
\end{equation}

The coefficient matrices are
\begin{equation}
C^{f}_{ij}=1+e^{i\phi_f}\epsilon^{\Delta^{f}_{ij}}+e^{i\psi_f}\epsilon^{\Delta^{\prime f}_{ij}}
\end{equation}
with $\epsilon=1/B$. The benchmark matrices imply the explicit $B$-scaling. Note that this model contains no texture zeros: every entry of $p^{f}_{ij}$ is finite, so every Yukawa entry $(Y_f)_{ij}$ is nonzero, and the observed hierarchy arises entirely from the differing rational powers of $\epsilon=1/B$ rather than from any vanishing entries (the $0$ entries denote an unsuppressed $\epsilon^{0}=1$ exponent, not a vanishing element; a texture zero would correspond to an infinite exponent).

\subsection{Matrix forms of the shifts \texorpdfstring{$\Delta,\Delta'$}{Delta, Delta-prime}}
The benchmark shifts retain the factorized form
$\Delta^f_{ij}=\alpha^f_i+\beta^f_j$,
$\Delta^{\prime f}_{ij}=\alpha^{\prime f}_i+\beta^{\prime f}_j$,
with the shift vectors (in units of $1/9$)
\begin{equation}
\begin{aligned}
\alpha^{u} &= \tfrac{1}{9}(2,4,1), &\quad \beta^{u} &= \tfrac{1}{9}(8,2,0),\\
\alpha^{\prime u} &= \tfrac{1}{9}(1,3,7), &\quad \beta^{\prime u} &= \tfrac{1}{9}(0,9,2),\\
\alpha^{d} &= \tfrac{1}{9}(5,1,1), &\quad \beta^{d} &= \tfrac{1}{9}(9,8,1),\\
\alpha^{\prime d} &= \tfrac{1}{9}(8,5,1), &\quad \beta^{\prime d} &= \tfrac{1}{9}(9,2,3),
\end{aligned}
\label{eq:shift-vectors}
\end{equation}
so that the $3\times 3$ shift matrices are
\begin{align}
\Delta^{u} &= \frac{1}{9}
\begin{pmatrix}
10 & 4 & 2\\
12 & 6 & 4\\
9 & 3 & 1
\end{pmatrix},
&
\Delta^{\prime u} &= \frac{1}{9}
\begin{pmatrix}
1 & 10 & 3\\
3 & 12 & 5\\
7 & 16 & 9
\end{pmatrix},
\label{eq:Delta-mats-u}
\\[4pt]
\Delta^{d} &= \frac{1}{9}
\begin{pmatrix}
14 & 13 & 6\\
10 & 9 & 2\\
10 & 9 & 2
\end{pmatrix},
&
\Delta^{\prime d} &= \frac{1}{9}
\begin{pmatrix}
17 & 10 & 11\\
14 & 7 & 8\\
10 & 3 & 4
\end{pmatrix}.
\label{eq:Delta-mats-d}
\end{align}
All shift entries are non-negative integers on the ninths lattice.
The four phases $\phi_u,\psi_u,\phi_d,\psi_d$ in \eqref{eq:fnmessenger} are
given in Eq.~\eqref{eq:app-phases}; together with the shifts of
Eq.~\eqref{eq:shift-vectors} they are determined by a fit to the quark masses
and the CKM magnitudes. The resulting effective coefficients satisfy
$0.42\le|C^u_{ij}|\le 1.60$ and $0.30\le|C^d_{ij}|\le 0.91$, so every entry is
of order unity, as required by the unit-magnitude messenger-chain
construction of Eq.~\eqref{eq:fnmessenger}.

\subsection{Reproducible diagonalization procedure}
\label{app:diag-procedure}

For reproducibility, we specify the diagonalization conventions
completely. Each Yukawa matrix is decomposed by singular value
decomposition as $Y_f=U_{f_L}\,\Sigma_f\,U_{f_R}^\dagger$ with
$\Sigma_f$ real, non-negative, and diagonal, so that
$Y_f^{\rm diag}=U_{f_L}^\dagger Y_f U_{f_R}$ as in
Eq.~\eqref{eq:bi-unitary}. Standard numerical routines return the
singular values in \emph{descending} order; the physical generation
assignment requires the \emph{ascending} order, so the singular values
and the corresponding columns of $U_{f_L}$ (and $U_{f_R}$) must be
reversed before forming
$V_{\rm CKM}=U_{u_L}^\dagger U_{d_L}$, placing the smallest singular
value in the $(1,1)$ position: generations $(u,c,t)$ and $(d,s,b)$
label the columns of $U_{f_L}$ in order of increasing mass. With
$\epsilon=14/75$, the exponent matrices of
Eq.~\eqref{eq:exponent-matrices}, the shift matrices of
Eqs.~\eqref{eq:Delta-mats-u}--\eqref{eq:Delta-mats-d}, and the
phases of Eq.~\eqref{eq:app-phases}, this procedure reproduces every
entry of Table~\ref{tab:benchmark-forward}: the mass ratios to the
quoted digits and all nine CKM magnitudes to the five quoted decimal
places, with $|J|=3.08\times10^{-5}$. Retaining the descending order
and reading the resulting matrix as though it were in conventional
generation order instead yields the generation-reversed matrix, with
the $(1,2)$ entry equal to $|V_{ts}|\simeq0.041$; no quantity quoted
in this paper is derived from that misordered reading. The same
ascending-order extraction is used throughout the Monte Carlo scan of
Sec.~\ref{sec:coefficient-sensitivity}.

\section{\texorpdfstring{$B$}{B}-Scalings of Wolfenstein Parameters}
\label{app:wolfenstein-scaling}

At leading Wolfenstein order define
\begin{equation}
\lambda \equiv |V_{us}|,\quad
A \equiv \frac{|V_{cb}|}{\lambda^{2}},\quad
r_{u} \equiv \frac{|V_{ub}|}{A\lambda^{3}},\quad
r_{t} \equiv \frac{|V_{td}|}{A\lambda^{3}}.
\end{equation}
In our $2/2$ scheme, with $\e=1/B$,
\begin{equation}
\lambda = \e^{8/9},\qquad
A = \e^{1/9},\qquad
A\lambda^{3} = \e^{25/9}.
\end{equation}
It follows that
\begin{equation}
r_{u} = \e^{5/9} = 0.3936,\qquad
r_{t} = \e^{1/18} = 0.9110.
\end{equation}
Solving the unitarity-triangle relations
\begin{equation}
r_{u}^{2} = \rho^{2}+\eta^{2},\qquad
r_{t}^{2} = (1-\rho)^{2}+\eta^{2}
\end{equation}
gives
\begin{equation}
\rho = \frac{1+r_{u}^{2}-r_{t}^{2}}{2} = 0.1625,\qquad
\eta = \sqrt{r_{u}^{2}-\rho^{2}} = 0.3585.
\end{equation}

\section{Algebraic Dictionary Between FX and PDG CKM Parameters}
\label{app:algebraic-dictionary}

\newcommand{\sU}{s_u}
\newcommand{\cU}{c_u}
\newcommand{\sD}{s_d}
\newcommand{\cD}{c_d}
\newcommand{\sT}{s}
\newcommand{\cT}{c}

\newcommand{\Vus}{|V_{us}|}
\newcommand{\Vub}{|V_{ub}|}
\newcommand{\Vcb}{|V_{cb}|}
\newcommand{\Vtd}{|V_{td}|}
\newcommand{\Vts}{|V_{ts}|}

We give a compact, algebraic dictionary between the Fritzsch--Xing (FX) and PDG conventions for the CKM matrix using rephasing-invariant inputs (four magnitudes and the Jarlskog invariant).

\subsection{Definitions}

Define
\begin{equation}
\begin{aligned}
\sU&=\sin\theta_u,\quad & \cU&=\cos\theta_u,\\
\sD&=\sin\theta_d,\quad & \cD&=\cos\theta_d,\\
\sT&=\sin\theta,\quad   & \cT&=\cos\theta.
\end{aligned}
\end{equation}
and the CP phase is $\phi_{\rm FX}$. In the PDG convention $s_{ij}=\sin\theta_{ij}$, $c_{ij}=\cos\theta_{ij}$ with CP phase $\delta\equiv\delta_{\rm PDG}$. We then define the FX matrix $V_{\rm CKM}^{\rm (FX)}$ as

\begingroup\footnotesize
\begin{equation}
\begin{pmatrix}
\sU\sD\,\cT+\cU\cD\,e^{-i\phi_{\rm FX}} &
\sU\cD\,\cT-\cU\sD\,e^{-i\phi_{\rm FX}} &
\sU\,\sT \\
\cU\sD\,\cT-\sU\cD\,e^{-i\phi_{\rm FX}} &
\cU\cD\,\cT+\sU\sD\,e^{-i\phi_{\rm FX}} &
\cU\,\sT \\
-\sD\,\sT &
-\cD\,\sT &
\cT
\end{pmatrix}
\label{eq:VFX}
\end{equation}
\endgroup

and the PDG matrix $V_{\rm CKM}^{\rm (PDG)}$ as
\begingroup\footnotesize
\begin{equation}
\begin{pmatrix}
c_{12}c_{13} &
s_{12}c_{13} &
s_{13}e^{-i\delta} \\
-s_{12}c_{23}-c_{12}s_{23}s_{13}e^{i\delta} &
c_{12}c_{23}-s_{12}s_{23}s_{13}e^{i\delta} &
s_{23}c_{13} \\
s_{12}s_{23}-c_{12}c_{23}s_{13}e^{i\delta} &
-c_{12}s_{23}-s_{12}c_{23}s_{13}e^{i\delta} &
c_{23}c_{13}
\end{pmatrix}
\label{eq:VPDG}
\end{equation}
\endgroup

\subsection{Invariant quantities}

From Eq.~\eqref{eq:VFX},
\begin{equation}
\begin{aligned}
\Vub&=\sU\sT,\qquad & \Vcb&=\cU\sT,\\
\Vtd&=\sD\sT,\qquad & \Vts&=\cD\sT.
\end{aligned}
\end{equation}
and
\begin{equation}
\Vus\simeq\left|\sU\cD\cT-\cU\sD e^{-i\phi_{\rm FX}}\right|.
\end{equation}

The Jarlskog invariant is
\begin{equation}
J=
\sU\cU\sD\cD\,\sT^{2}\cT\sin\phi_{\rm FX}
=
s_{12}s_{23}s_{13}\,c_{12}c_{23}c_{13}^{2}\sin\delta.
\end{equation}

\subsection{Compact dictionary (PDG \texorpdfstring{$\to$}{to} FX)}

Given PDG magnitudes $\{\Vus,\Vub,\Vcb,\Vtd\}$ and $\delta$:
\begin{align}
\sT &= \sqrt{\Vub^{2}+\Vcb^{2}}, &
\theta_u &= \arctan\!\left(\frac{\Vub}{\Vcb}\right), \nonumber\\
\Vts &\simeq \sqrt{\Vub^{2}+\Vcb^{2}-\Vtd^{2}}, &
\theta_d &= \arctan\!\left(\frac{\Vtd}{\Vts}\right).
\end{align}

then $\cos\phi_{\rm{FX}}$ and $\sin\phi_{\rm{FX}}$ are given by:

\begin{equation}
\cos\phi_{\rm FX}=
\frac{\sU^{2}\cD^{2}\cT^{2}+\cU^{2}\sD^{2}-\Vus^{2}}
{2\,\sU\cU\sD\cD\cT}
\label{eq:cosphiFX-dict-app}
\end{equation}

\begin{equation}
\sin\phi_{\rm FX}=
\frac{J}{\sU\cU\sD\cD\,\sT^{2}\cT}.
\label{eq:sinphiFX-dict-app}
\end{equation}

\noindent The physically meaningful equality is the equality of $J$, not the numerical equality of phase angles across conventions.

\clearpage


\end{document}